\documentclass[10pt]{article}

\usepackage{cite}
\usepackage{times}
\usepackage{graphicx}

\title{Dynamical decoupling and noise spectroscopy\\ with a superconducting flux qubit\\}

\author{Jonas Bylander$^{1}$,
Simon Gustavsson$^{1}$,\\
Fei Yan$^{2}$,
Fumiki Yoshihara$^{3}$,
Khalil Harrabi$^{3\,\dagger}$,
George Fitch$^{4}$,\\
David G.\@ Cory$^{2,5,6}$,
Yasunobu Nakamura$^{3,7}$,
Jaw-Shen Tsai$^{3,7}$,
William D.\@ Oliver$^{1,4}$
\\~\\
 \normalsize{$^{1}$Research Laboratory of Electronics,
 $^{2}$Department of Nuclear Science and Engineering,} \\
 \normalsize{Massachusetts Institute of Technology, Cambridge, MA 02139, USA.}\\
 \normalsize{$^{3}$The Institute of Physical and Chemical Research (RIKEN), Wako, Saitama 351-0198, Japan.}\\
 \normalsize{$^{4}$MIT Lincoln Laboratory, 244 Wood Street, Lexington, MA 02420, USA.}\\
 \normalsize{$^{5}$Institute for Quantum Computing and Dept.\@ of Chemistry, University of Waterloo, ON, N2L 3G1, Canada.} \\
 \normalsize{$^{6}$Perimeter Institute for Theoretical Physics, Waterloo, ON, N2J, 2W9, Canada.} \\
 \normalsize{$^{7}$Green Innovation Research Laboratories, NEC Corporation, Tsukuba, Ibaraki 305-8501, Japan.}\\
 \normalsize{$^\dagger$Present address: Physics Dept.\@, King Fahd University of Petroleum \& Minerals, Dhahran 31261, Saudi Arabia.}
 }

\date{\today}

\begin{document}


\maketitle

\subsection*{Abstract}
  The characterization and mitigation of decoherence in natural and artificial two-level systems (qubits) is fundamental to quantum information science and its applications.
  Decoherence of a quantum superposition state arises from the interaction between the constituent system and the uncontrolled degrees of freedom in its environment.
  Within the standard Bloch-Redfield picture of two-level system dynamics, qubit decoherence is characterized by two rates:
  a longitudinal relaxation rate $\Gamma_1$ due to the exchange of energy with the environment, and a transverse relaxation rate $\Gamma_2 = \Gamma_1/2 + \Gamma_\varphi$ which contains the pure dephasing rate $\Gamma_\varphi$.
  Irreversible energy relaxation can only be mitigated by reducing the amount of environmental noise, reducing the qubit's internal sensitivity to that noise, or through multi-qubit encoding and error correction protocols (which already presume ultra-low error rates).
  In contrast, dephasing is in principle reversible and can be refocused dynamically through the application of coherent control pulse methods\cite{Hahn-PR-50,Slichter-NatureStyleRef,Viola-PRA-98}.
  In this work we demonstrate how dynamical-decoupling techniques can moderate the dephasing effects of low-frequency noise on a superconducting qubit\cite{Mooij-Science-99,Orlando-PRB-99,Clarke-Nature-08} with energy-relaxation time $T_1 = 1/\Gamma_1 = 12\,\mu$s.
  Using the CPMG sequence\cite{Carr-Purcell-PR-54,Meiboom-Gill-RSI-58} with up to 200 $\pi$-pulses, we demonstrate a 50-fold improvement in the transverse relaxation time $T_2$ over its baseline value.
  We observe relaxation-limited times $T_2 ^{\mathrm{\,CPMG}} = 23 \,\mu \textrm{s} \approx 2\,T_1$ resulting from CPMG-mediated Gaussian pure-dephasing times in apparent excess of $100 \,\mu \textrm{s}$.
  We leverage the filtering property of this sequence in conjunction with Rabi and energy relaxation measurements
  to facilitate the spectroscopy and reconstruction of the environmental noise power spectral density (PSD)~\cite{Clerk-RMP-10,Astafiev-PRL-04}.

\clearpage

Several multi-pulse sequences developed within the field of nuclear magnetic resonance\cite{Slichter-NatureStyleRef} (NMR) have recently been applied to mitigate noise in qubits based on atomic ensembles\cite{Biercuk-Nature-09}, semiconductor quantum dots\cite{Bluhm-arxiv-10,Barthel-PRL-10}, and diamond nitrogen--vacancy centres\cite{deLange-Science-10,Ryan-PRL-10}.
We extend these methods to the realm of superconducting quantum devices, and subject a remarkably long-lived qubit to varying levels of longitudinal and transverse noise by rotating the qubit's quantization axis, against which we characterize the baseline coherence rates $\Gamma_1$, $\Gamma_2$, and $\Gamma_\varphi$.
We evaluate three dynamical-decoupling pulse protocols: the Carr-Purcell\cite{Carr-Purcell-PR-54} (CP); Carr-Purcell-Meiboom-Gill\cite{Meiboom-Gill-RSI-58} (CPMG); and Uhrig dynamical-decoupling\cite{Uhrig-PRL-07} (UDD) sequences.
The narrow-band filtering property of the CPMG sequence enables us to sample environmental noise over a broad frequency range 0.2--20\,MHz, and we observe a $1/f^{\alpha}$-type spectrum which we independently confirm with a Rabi-spectroscopy approach.
We furthermore characterize the environmental noise from 5.4 to 21\,GHz by monitoring the qubit's relaxation rate\cite{Clerk-RMP-10,Astafiev-PRL-04}.

\begin{figure}[t!]
\begin{center}
\includegraphics[width=15cm]{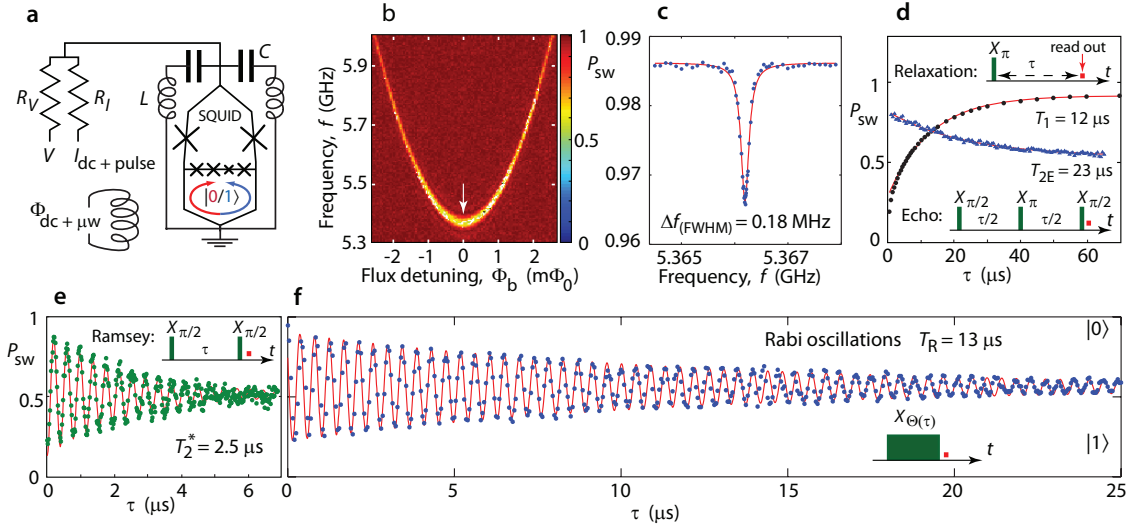}
\vspace{-0.5cm}
\begin{quote}
\caption{\label{fig:time_domain} \textbf{Qubit device and characterization.}
\textbf{a}, Device and biasing schematic: An aluminium superconducting loop interrupted by Josephson junctions (crosses) with a read-out dc SQUID.
    \textbf{b}, Frequency spectroscopy of the qubit's $|0\rangle\to|1\rangle$ transition.
    \textbf{c}, Spectroscopy at $\Phi_\mathrm{b}=0$ (arrow in \textbf{b}).
    \textbf{d}, Echo decay (blue triangles) and relaxation from the excited state (black dots) at $\Phi_\mathrm{b}=0$.
    In the insets, $\tau$ is a time delay and $X_\Theta$ symbolizes a rotation of the Bloch vector by the angle $\Theta$ around the axis $\hat{\sigma}_x$.
    The red squares indicate the time of read out.
    \textbf{e}, Free-induction decay (Ramsey fringe) at $\Phi_\mathrm{b}=0$.
    \textbf{f}, Rabi oscillations at $\Phi_\mathrm{b}=0$.
    }
\end{quote}
\vspace{-1cm}
\end{center}
\end{figure}

The device is a persistent-current qubit (Figs.~\ref{fig:time_domain}a and A1), an aluminium loop interrupted by four Al-AlO$_x$-Al Josephson junctions.
When an external magnetic flux $\Phi$ threading the loop is close to half a superconducting flux quantum $\Phi_0/2$,
the diabatic states correspond to clockwise (counterclockwise) persistent currents $I_\mathrm{p}=0.18\,\mu$A with energies $\pm \hbar \varepsilon/2=\pm I_\mathrm{p} \Phi_{\mathrm{b}}$, tunable by the flux bias $\Phi_{\mathrm{b}}=\Phi-\Phi_0/2$.
At $\Phi_{\mathrm{b}}=0$, the degenerate persistent-current states hybridize with a strength $ \hbar \Delta = h\times 5.3662$~GHz
(Figs.~\ref{fig:time_domain}b and A2b), where $\hbar = h/2 \pi$ and $h$ is Planck's constant.
The corresponding two-level Hamiltonian is\cite{Orlando-PRB-99,Mooij-Science-99}
\begin{equation}
    \hat{\mathcal{H}} = - \frac{\hbar}{2} \left[ (\varepsilon + \delta \varepsilon) \hat{\sigma}_{x} + (\Delta + \delta \Delta) \hat{\sigma}_{z} \right],
    \label{eq:Hamiltonian}
\end{equation}
which includes noise fluctuation terms $\delta \varepsilon$ and $\delta \Delta$, and $\hat{\sigma}_{x,z}$ are Pauli operators (Fig.~A2).
The ground $\left( |0\rangle \right)$ and excited $\left( |1\rangle \right)$ states have frequency splitting $\omega_{01} = \sqrt{\varepsilon^2+\Delta^2}$ and are well isolated
owing to the qubit's large anharmonicity, $\omega_{12}/\omega_{01}\approx5$.
The environmental noise leading to fluctuations $\delta \varepsilon$ (\emph{e.g.\@}, flux noise) and $\delta \Delta$ (\emph{e.g.\@}, critical current and charge noise) physically couples to the qubit in the $\varepsilon$ -- $\Delta$ frame (equation~\ref{eq:Hamiltonian}).
However, their manifestation as longitudinal noise (dephasing) or transverse noise (energy relaxation) is tunable\cite{Ithier-PRB-05} by the flux bias $\Phi_{\mathrm{b}}$ and determined, respectively, by their projections $\delta \omega_{z'}$ onto the qubit's quantization axis $\hat{\sigma}_{z'}$ (which makes an angle $\theta=\arctan(\varepsilon/\Delta)$ with $\hat{\sigma}_{z}$) and $\delta \omega_{\perp '}$ onto the plane perpendicular to $\hat{\sigma}_{z'}$.

The chip is mounted in a $^3\textrm{He}/^4\textrm{He}$ dilution refrigerator with 12-mK base temperature.
For each experimental trial, we initialize the qubit by waiting sufficient time ($\sim\!1$~ms) for it to relax to its ground state.
We drive the desired quantum-state rotations of angle $\Theta$ by applying calibrated in-phase ($X_{\Theta}$) and quadrature ($Y_{\Theta}$) harmonic flux pulses to the qubit loop.
The pulses comprise Gaussian envelopes with a typical standard deviation $\sigma=1.2$\,ns and truncated at $\pm3\sigma$.
The qubit readout has 79\% visibility (Fig.~A3) and is performed in the energy basis by determining the switching probability $P_\mathrm{sw}$ of a hysteretic dc SQUID, averaging over several thousand trials
(see Appendix).

\begin{figure}[b!]
\begin{center}
\includegraphics[width=12cm]{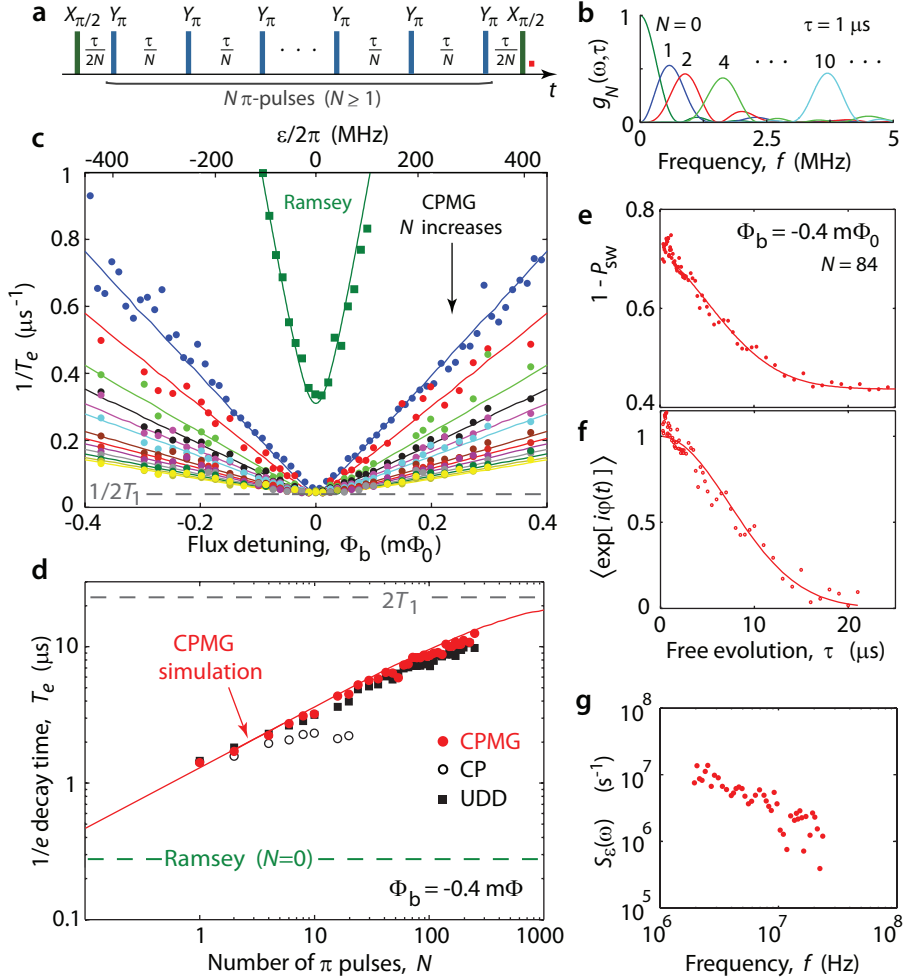}
\vspace{-1cm}
\begin{quote}
\caption{\label{fig:cpmg} \textbf{Dynamical decoupling pulse sequences.}
\textbf{a--b}, CPMG pulse sequence and filter function $g_N(\omega,\tau)$ for a fixed total pulse-sequence length $\tau$. 
    \textbf{c}, Decay rates (inverse of $1/e$ times) vs.\@ flux detuning:
    Free induction (green squares) and CPMG (coloured dots) with $N=1,2,4,6,8,10,16,20,24,30,36,42,48$. Solid lines are calculations using equation~(\ref{eq:chi}) with parameters in Table 1.
    \textbf{d--g}, Measurements at $\Phi_\mathrm{b}=-0.4\,\mathrm{m}\Phi_0$  ($\varepsilon/2\pi=430$\,MHz).
    \textbf{d}, $1/e$ decay time under $N$-pulse CPMG, CP, and UDD sequences. The simulation (red line) assumes perfect pulses.
    \textbf{e}, Population decay under an 84-pulse CPMG sequence.
    Gaussian fit with fixed exponential decay contributions
    $T_1=12\,\mu$s, $T_\mathrm{p}=1.75\,\mu$s.
    \textbf{f}, Phase decay of the signal in \textbf{e} after dividing out $T_1$ and $T_\mathrm{p}$.
    \textbf{g}, Noise PSD calculated from the data in \textbf{f}.
    }
   \end{quote}
\end{center}
\end{figure}

\begin{figure}[b!]
\begin{center}
\includegraphics[width=6cm]{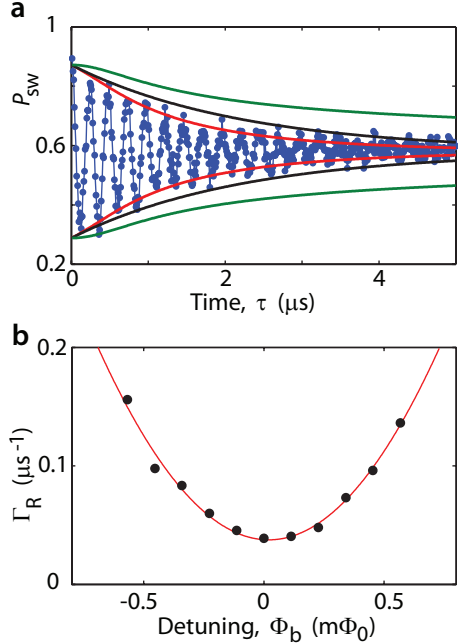}
\vspace{-0.5cm}
\begin{quote}
\caption{\label{fig:Rabi_with_Sim} \textbf{Decoherence during driven dynamics.}
\textbf{a}, Rabi oscillations with $\Omega_\mathrm{R}/2\pi=2$\,MHz at $\varepsilon/2\pi=225$\,MHz.
    The red line envelope is a fitting using $\zeta(t)$ and $\Gamma_\mathrm{R}$.
    The black line shows the $\Gamma_\mathrm{R}$ decay only, and the green line the $\zeta(t)$-envelope contribution.
    \textbf{b}, Rabi-decay rate $\Gamma_\mathrm{R}$ vs. flux detuning at $\Omega_\mathrm{R}/2\pi=2$\,MHz, with a parabolic fit to equation~(\ref{eq:Gamma_R_simplified})
    used to obtain $S_\varepsilon(\Omega_\mathrm{R})=\frac{1}{\pi}\,\Gamma_\Omega^{(\varepsilon)}$.
    }
\end{quote}
\end{center}
\end{figure}

We begin with a spectroscopic characterization of our device and its baseline coherence times.
The qubit level splitting $\omega_{01}$ is measured via saturated frequency spectroscopy (Fig.~\ref{fig:time_domain}b), and at low-power it exhibits a Lorentzian full-width-at-half-maximum (FWHM) linewidth $\Delta f_\mathrm{(FWHM)} = 0.18$\,MHz at $\Phi_{\mathrm{b}}=0$ (Fig.~\ref{fig:time_domain}c).
The energy relaxation is generally exponential and its time constant $T_1=12\pm1\,\mu$s (Fig.~\ref{fig:time_domain}d) is remarkably long among superconducting qubits\cite{Clarke-Nature-08},
a feature we leverage in this work in conjunction with quantization axis tunability.
We observe similarly long decay times at $\Phi_{\mathrm{b}}=0$ for the Hahn spin-echo, $T_{2,\mathrm{E}}=23\,\mu$s (Fig.~\ref{fig:time_domain}d), Ramsey free induction, $T_2^{^*}=2.5\,\mu$s (Fig.~\ref{fig:time_domain}e), and Rabi oscillations, $T_\mathrm{R}=13\,\mu$s (Fig.~\ref{fig:time_domain}f).
Although the spin echo and Rabi exhibit an apparently exponential decay at $\Phi_{\mathrm{b}}=0$ (they are essentially $T_1$-limited at this flux bias), in general, their decay functions are non-exponential.
Furthermore, the dephasing times decrease (rates increase) dramatically away from $\Phi_{\mathrm{b}}=0$ (Figs.~\ref{fig:cpmg}c and~\ref{fig:Rabi_with_Sim}b) due to the qubit's increased sensitivity to the dominant $\delta \varepsilon$-noise (flux noise) in this system\cite{Yoshihara-PRL-06} (increased $|\partial\omega_{01}/\partial\varepsilon|$), which, as we will demonstrate, can be mitigated with multi-pulse dynamical-decoupling sequences to push coherence times up towards the $T_1$-limit.

Decoherence in superconducting qubits has been studied theoretically\cite{Averin-ForPh-00,Makhlin-RMP-01,Ithier-PRB-05,Clerk-RMP-10} and experi\-men\-tally\cite{Astafiev-PRL-04,Ithier-PRB-05,Yoshihara-PRL-06}.
Each noise source $\lambda$ is characterized by its PSD,
which quantifies the frequency distribution of the noise power, $S_\lambda(\omega) = (1/2\pi) \int_{-\infty}^\infty \mathrm{d}t \, \langle \lambda(0)\lambda(t) \rangle \exp(-i\omega t)$.
In contrast to a simple Bloch-Redfield picture, which presumes weakly coupled noise sources with short correlation times $\tau_{\mathrm{c}} \ll T_1,T_2$ resulting in purely exponential decay functions,
more generally the decay functions are non-exponential for noise sources with long correlation times, singular near $\omega \approx 0$ (\emph{e.g.\@}, $1/f$-type noise at low frequencies, relevant to dephasing).

A superposition state's accumulated phase $\varphi(t) = \langle\omega_{01}\rangle t + \delta\varphi(t)$
diffuses due to adiabatic fluctuations of the transition frequency, $\delta\varphi(t) = (\partial\omega_{01}/\partial\lambda) \int_0^t \mathrm{d}t'\delta\lambda(t')$.
For noise generated by a large number of fluctuators that are weakly coupled to the qubit, its statistics are Gaussian.
Ensemble averaging over all realizations of the stochastic process $\delta\lambda(t)$, and taking the sources $\lambda$ to be independent, the dephasing is
$\langle \exp[i\,\delta\varphi(t)] \rangle \equiv \exp[-\chi_N(t)]$, with the coherence integral
\begin{equation} \label{eq:chi}
 \chi_N(\tau) = \tau^2 \sum_{\lambda} \left(\frac{\partial\omega_{01}}{\partial\lambda}\right)^2 \int_0^\infty \mathrm{d}\omega \, S_\lambda(\omega) \, g_N(\omega,\tau) ,
\end{equation}
where $\tau$ is the free evolution time and $N$ will denote the number of $\pi$ pulses in the pulse sequences\cite{Uhrig-PRL-07,Biercuk-Nature-09}.
Equation~(\ref{eq:chi}) expresses that the PSD $S_\lambda(\omega)$ is filtered by a dimensionless weighting function $g_N$ determined by the pulse sequence,
and the aggregated $\lambda$-noise translates to
dephasing through $\partial\omega_{01} / \partial\lambda$, the qubit's longitudinal sensitivity to $\lambda$-noise.
For Gaussian noise with spectral distribution $S_{\lambda}(\omega) = A_{\lambda}/\omega$ at low frequencies, the coherence integral results in a Gaussian decay function, $\chi_N(\tau) = (\Gamma_{\varphi} \tau)^2$.

%
The Ramsey free-induction and Hahn spin-echo dependence on flux bias (Fig.~\ref{fig:cpmg}c) are both apparently consistent with Gaussian distributed, $1/f$-type noise.
Ramsey free induction, the free evolution of a superposition state for a time $\tau$ (Fig.~\ref{fig:cpmg}a with no $\pi$ pulses), has a filter function $g_0$
peaked at $\omega=0$ (Fig.~\ref{fig:cpmg}b) and is sensitive to low-frequency longitudinal noise $\delta\omega_{z'}(\omega \rightarrow 0)$.
Inhomogeneities in the precession frequency $\omega_{01}$ from one realization of the pulse sequence to the next lead to a decay of the averaged signal.
We denote such fluctuations ``quasi-static" 
noise and characterize them by a noise variance $\sigma_\lambda^2 = 2 \int_{\omega^\lambda_\mathrm{ir}}^{\omega^\lambda_\mathrm{uv}} \, \mathrm{d}\omega \, S_\lambda(\omega)$,
with cut-off frequencies
$\omega^\lambda_\mathrm{ir}$ and $\omega^\lambda_\mathrm{uv}$ determined respectively by the averaging time over all trials and the typical free-evolution time during a single trial.
In contrast, the Hahn spin-echo sequence\cite{Hahn-PR-50}, a single $\pi$ pulse applied at time $\tau/2$ (Fig.~\ref{fig:cpmg}a with one $\pi$ pulse), has a filter function $g_1$ peaked away from $\omega=0$ (Fig.~\ref{fig:cpmg}b) and is less sensitive to quasi-static noise.
We plot the decay rates $1/T_e$ versus flux bias $\Phi_{\mathrm{b}}$ for Ramsey and echo in Fig.~\ref{fig:cpmg}c (see also Fig.~A4), where, for purposes of comparison amongst different decay envelopes, $T_e$ parameterizes the time $T_2$ to decay by a factor $1/e$ independent of the exact decay function (see Appendix).
At $\Phi_{\mathrm{b}}=0$, $\delta \Delta$-noise is the dominant longitudinal noise that limits the Ramsey decay, yet it is refocused with a single $\pi$-pulse resulting in the $T_1$-limited  exponential echo-decay in Fig.~\ref{fig:time_domain}d.
As $|\Phi_{\mathrm{b}}|$ is increased, both the Ramsey and echo decay rates increase due to the qubit's increased longitudinal sensitivity to $\delta \varepsilon$-noise (see Figs.~\ref{fig:cpmg}c and A4).
The $\delta \varepsilon$-noise is too large for the echo to refocus efficiently, due to its high-frequency tail, and we find best-fit phase decay functions that are Gaussian,
$\chi(\tau) = (\Gamma_{\varphi,\mathrm{F(E)}}\,\tau)^2$.
We extract the ratio $\Gamma_{\varphi,\mathrm{F}}(\Phi_{\mathrm{b}}) / \Gamma_{\varphi,\mathrm{E}}(\Phi_{\mathrm{b}})  \approx 4.5$,
as expected for $1/f$ noise\cite{Ithier-PRB-05}, with an equivalent flux-noise amplitude\cite{Yoshihara-PRL-06}
$A_\Phi = (1.7\,\mu\Phi_0)^2$.
Importantly, we note that in this analysis and related works\cite{Ithier-PRB-05,Yoshihara-PRL-06}, the PSD was presumed {\em a priori \em} to take the form $1/f^{\alpha}$ with $\alpha=1$.

Numerical simulations (see Appendix), including the measured $T_1$ decay at each $\Phi_{\mathrm{b}}$, linearly coupled and uncorrelated quasi-static noises, and uncorrelated dynamic $1/f$ noise from $10^4$ to $10^{10}$  Hz, reproduce the entire $\Phi_{\mathrm{b}}$-dependence of $T_2^{^*}$ and $T_{2,\mathrm{E}}$ using the parameters in Table~1 and are also consistent with equation~(\ref{eq:chi}).

In order to mitigate higher-frequency noise than what the Hahn spin-echo can efficiently refocus, we further shape the filter function by applying additional $\pi$-pulses (Figs.~\ref{fig:cpmg}a and A5).
The filter $g_N(\omega,\tau)$ depends on the number $N$ and distribution of $\pi$-pulses\cite{Uhrig-PRL-07,Cywinski-PRB-08,Biercuk-Nature-09,Biercuk-PRA-09}
during the total sequence length $\tau$,  
\begin{equation} 
  g_N(\omega,\tau) =
  \frac{1}{(\omega\tau)^2} \, \Big| 1+(-1)^{1+N}\exp(i\omega\tau) + 2\sum_{j=1}^{N} (-1)^j \exp(i\omega\delta_j\tau) \cos(\omega\tau_\pi/2) \Big|^2 ,
\end{equation}
where $\delta_j \in [0,1]$ is the normalized position of the centre of the $j$th $\pi$-pulse between the two $\pi/2$-pulses, and $\tau_\pi$ is the length of each $\pi$-pulse.
As the number of pulses increases for fixed $\tau$, the filter function's peak shifts to higher frequencies (Fig.~\ref{fig:cpmg}b), leading to a reduction in the net integrated noise (equation~\ref{eq:chi}) for $1/f^{\alpha}$-type noise spectra with $\alpha > 0$.
Alternatively, for a fixed time separation $\tau' = \tau/N$ (valid for $N \ge 1$), the filter asymptotically peaks near $\omega' = 2\pi/4\tau'$ as more pulses are added.
In principle, one can adapt the filter function to suit a particular noise spectrum via the choice of dynamical-decoupling protocol.

We have evaluated three different dynamical-decoupling protocols relevant for $1/f$-type power law noise spectra\cite{Faoro-PRL-04,Falci-PRA-04,Pasini-PRA-10}.
The CP and CPMG sequences\cite{Carr-Purcell-PR-54,Meiboom-Gill-RSI-58} are multi-pulse extensions of the Hahn echo with equally spaced $\pi$-pulses whose phases differ from the initial $\pi/2$ pulse by $0^\circ$ ($X_\pi$) and $90^\circ$ ($Y_\pi$), respectively (Fig.~\ref{fig:cpmg}a).
The UDD sequence\cite{Uhrig-PRL-07} 
has $Y_\pi$-pulse positions defined by
$\delta_j = \sin^2 \left( \frac{\pi j}{2N+2} \right)$.

In Fig.~\ref{fig:cpmg}c, we include the $1/T_e$ decay rates for CPMG dynamical-decoupling sequences with $N=2 \ldots 48$ $\pi$-pulses along with the Ramsey ($N=0$ ) and Hahn echo ($N=1$) already discussed.
The decay rates monotonically improve towards the $1/2T_1$-limit as the number of $\pi$-pulses increases, extending the range around $\Phi_{\mathrm{b}}=0$ for which dephasing is negligible.
Outside this range, as the qubit's sensitivity to $\delta \varepsilon$ (flux) noise grows with increasing $|\Phi_\mathrm{b}|$, increasing the number of $\pi$-pulses mitigates the noise sufficiently well to achieve a desired decay rate.

At a specific flux bias $\Phi_{\mathrm{b}} = 0.4 \; \textrm{m}\Phi_0$, where the qubit is highly sensitive to $\delta \varepsilon$ noise, the CPMG sequence gives a marked improvement in the decay time $T_e$ up to $N\approx 200$ $\pi$-pulses (Fig.~\ref{fig:cpmg}d), beyond which pulse errors begin to limit the CPMG efficiency.
We achieve a 50-fold enhancement of $T_2^\mathrm{\,CPMG}$ over the Ramsey $T_2^{^*}$ (Fig.~\ref{fig:cpmg}d), and well over 100-fold improvement in the Gaussian pure dephasing time $T_{\varphi}$.
The CPMG sequence performs about 5\% better than UDD, indicating that the $1/f$ $\delta\varepsilon$-noise spectrum exhibits a relatively soft (if any) ultraviolet cutoff\cite{Cywinski-PRB-08,Pasini-PRA-10},
and it dramatically outperforms CP, as $Y_\pi$-pulse errors appear only to fourth order with CPMG, whereas with CP, $X_\pi$ errors accumulate to second order\cite{Borneman-JMR-10}.

\begin{figure}[b!]
\begin{center}
\includegraphics[width=14cm]{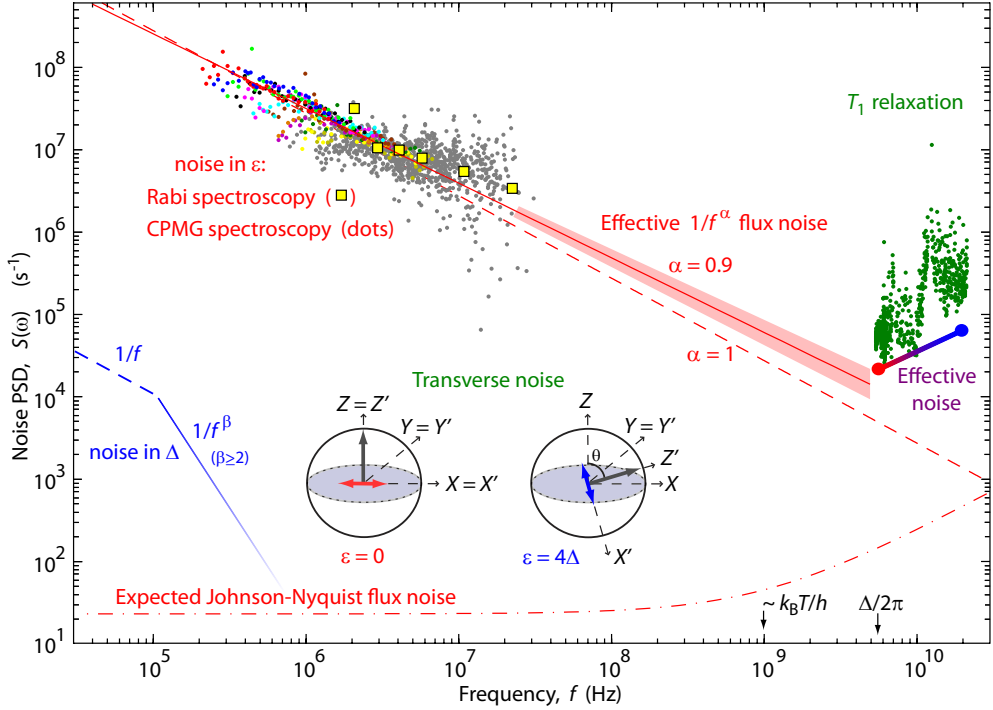}
\begin{quote}
\caption{\label{fig:psd} \textbf{Noise-power spectral density (PSD).}
\textbf{Multi-coloured dots}, $\delta\varepsilon$-noise PSD (0.2--20\,MHz) derived from CPMG data at $\Phi_\mathrm{b}\! =\! -0.4\,\mathrm{m}\Phi_0$ (see text).
    Colours correspond to the various $N$ in Fig.~\ref{fig:cpmg}b--c; grey dots for data up to $N\!=\!250$.
    \textbf{Yellow squares}, $\delta\varepsilon$-noise PSD (2--20\,MHz) derived from Rabi spectroscopy (Fig.~\ref{fig:Rabi_with_Sim}; see text).
    \textbf{Diagonal, dashed lines}, Estimated $1/f$ flux (red) and $\delta\Delta$ (blue) noise inferred from the Ramsey and echo measurements (\emph{cf.\@} $A_\lambda$ and $\omega^\Delta_\mathrm{uv}$ in Table~1).
    \textbf{Solid, red line}, Power-law dependence, $S_\varepsilon(f) = A'_\varepsilon/(2\pi f)^\alpha$, extrapolated beyond the qubit's frequency, $\Delta/2\pi$.
    The parameters $A'_\varepsilon = (0.8\,\mu\Phi_0)^2$ and $\alpha=0.9$ were determined by fitting the low-frequency, linear portion of the CPMG PSD data before the slight upturn beyond 2 MHz (see Appendix).
    The shaded area covers $\alpha\pm0.05$.
    \textbf{Green dots}, High-frequency $\varepsilon$ and $\Delta$ PSD inferred from energy-relaxation measurements above $\Delta/2\pi\!=\!5.4$~GHz.
    \textbf{Purple line}, Guide to indicate linearly increasing Nyquist (quantum) noise, including the eigenbasis rotation (see inset);
    dots indicate transverse $\delta\varepsilon$ (red) and $\delta\Delta$ (blue) noises.
    \textbf{Inset},
    Graphic representation of the quantization axis (grey arrows of fixed length) with the qubit's ($Z'$) eigenstate tilted from the ``laboratory'' frame ($Z$) by the angle $\theta$.
    Fields $\varepsilon(\Phi_\mathrm{b})$ and $\Delta$ point in the $X$ and $Z$ directions, respectively.
    Red and blue double-arrows indicate transverse noise.
    }
\end{quote}
\end{center}
\end{figure}

We use the filtering property of the CPMG sequence to characterize the $\delta \varepsilon$-noise spectrum.
The filter $g_N(\omega,\tau)$ is sufficiently narrow about $\omega'$ that we can treat the noise as constant within its bandwidth $B$ and approximate equation~(\ref{eq:chi}) as
$\chi_N(\tau) \approx \tau^2 \, (\partial\omega_{01}/\partial\Phi)^2 \, S_\varepsilon(\omega') \, g_N(\omega' ,\tau') \, B$ (Fig.~A6).
We compute $\omega'$ and $B$ numerically for each $N$ and $\tau$ used in the CPMG measurements of Fig.~\ref{fig:cpmg}d.
The measured decay function contains three decay rates:
dephasing $\Gamma_\varphi^{(N)}$ and exponential relaxation $\Gamma_1/2$ during the total free-evolution time $\tau$,
and pulse-induced decay $\Gamma_\mathrm{p}$ during $N\tau_\pi$ (we assume $\Gamma_\mathrm{p}$ to be independent of $N$).
Conceptually, Figs.~\ref{fig:cpmg}e--g illustrate for $N=84$ how we divide out the $\Gamma_1$ and $\Gamma_\mathrm{p}$ (assuming exponential pulse-induced decay) components from the raw data (e) to obtain the Gaussian phase decay (f), and then compute $S_\varepsilon(\omega)$ (g).
More rigorously, we only use that method to determine a starting point at a single frequency, and then use a recursive method to obtain the remainder of the spectrum without presuming a functional form for the pulse-induced decay, the dephasing, or the noise spectrum (see Appendix).
Both approaches yield nearly identical $1/f^{\alpha}$-type spectra with a slight increase in the measured PSD above 2 MHz;
%
we plot the recursively extracted PSD $S_{\varepsilon}(\omega)$ over the region 0.2--20\,MHz in Fig.~\ref{fig:psd}.
Interestingly, the PSD estimated in this manner is better approximated by a $1/f^{\alpha}$ power law\cite{Wellstood-APL-87} with $\alpha = 0.9 < 1$ (solid, red line) with noise amplitude $A_\Phi = (0.8\,\mu\Phi_0)^2$, obtained by fitting the lower-frequency, linear portion of the PSD.
Projecting this line to higher frequencies comes within a factor 2 of the transverse noise at frequency $\Delta$ as extracted from the relaxation measurements described below.


We confirmed the spectrum over a similar frequency range by analyzing the decoherence during driven evolution,
which provides an independent means to rotate the quantization axis with respect to the noise sources when viewed in the rotating frame\cite{Geva-Skinner-JChemPhys-95,Ithier-PRB-05}.
A transverse driving field at frequency $\omega$ results in Rabi oscillations with angular frequency
$\Omega_\mathrm{R} = \sqrt{\Omega^2 + (\Delta\omega)^2} \approx \Omega + (\Delta\omega)^2/2\Omega$,
where $\Delta\omega = \omega - \omega_{01}$.
Integrating the oscillations over a normal distribution with variance $\sigma_\varepsilon^2$, we obtain the quasi-static decay function $\zeta(\tau) = \left(1 + (u \tau)^2 \right)^{-1/4}$,
where
$u = (\varepsilon/\omega_{01})^2 \, \sigma_\varepsilon^2/\Omega$.
Along with $\Gamma_1$, the noise at the Rabi frequency $\Gamma_\Omega^{(\lambda)} = \pi S_{\lambda}(\Omega_\mathrm{R})$ comprises the usual exponential Rabi-decay rate
\begin{equation} \label{eq:Gamma_R_simplified}
 \Gamma_\mathrm{R} = \left( \frac{3}{4}\Gamma_1 + \frac{1}{2}\Gamma_\Omega^{(\Delta)} \right) \cos^2\theta + \left(\frac{\varepsilon}{\omega_{01}}\right)^2 \, \frac{1}{2}\Gamma_\Omega^{(\varepsilon)} ,
\end{equation}
where $\cos^2\theta\!\approx\!1$ as the quantization angle is small.
The combined decay function is $\zeta(\tau) \exp \left( -\Gamma_{\mathrm{R}} \tau\right)$ \, (Fig.~\ref{fig:Rabi_with_Sim}a; see also Appendix).

To determine $S_\lambda(\Omega_\mathrm{R})$, we measured the Rabi oscillations vs.\@ $\Phi_{\mathrm{b}}$ with fixed Rabi frequency $\Omega_\mathrm{R}$.
For each $\Omega_\mathrm{R}$, we find the $\varepsilon$-independent part of the rate by fitting the envelope of the oscillations at $\varepsilon=0$.
The rate $\Gamma_\Omega^{(\Delta)}$ was too small to distinguish accurately from $\Gamma_1$, consistent with its correspondingly small quasi-static noise $\sigma_{\Delta}^2$ (Table~1).
Then, for $\varepsilon \neq 0$, we divide out the known quasi-static contribution $\zeta(\tau)$ and fit to the parabolic term in equation~(\ref{eq:Gamma_R_simplified}), from which we obtain
$\Gamma_\Omega^{(\varepsilon)}$ (Fig.~\ref{fig:Rabi_with_Sim}b).
Using this approach, we find $S_\varepsilon(\Omega)$ to be consistent with the $1/f^{\alpha}$ noise obtained from the CPMG measurements (Fig.~\ref{fig:psd}).

\begin{table}[b!]
 \label{table:times}
 \begin{center}
 \caption{\textbf{Quasi-static noise parameters used in simulations, and coherence times.}}
 \vspace{0.5cm}
 \begin{tabular}{|c||c|c|c|c|}
    \hline
    \textbf{Noise parameters} & $\sigma_{\lambda}/2\pi$ & $\omega^\lambda_{\mathrm{ir}}/2\pi$ & $\omega^\lambda_{\mathrm{uv}}/2\pi$ & $A_{\lambda}$ \\
    \hline
    $\lambda = \varepsilon \,\,\mathrm{(equiv.} \,\,\Phi\, \mathrm{noise)}$      & $10\,\mathrm{MHz}$  & $1\,\mathrm{Hz}$ & $1\,\mathrm{MHz}$   & $(1.7 \times 10^{-6})^2 \; \Phi_0^2$ \\
    $\lambda = \Delta \,\,\mathrm{(equiv.} \,\,I/I_\mathrm{c}\, \mathrm{noise)}$ & $0.06\,\mathrm{MHz}$  & $1\,\mathrm{Hz}$ & $< 0.1\,\mathrm{MHz}$ & $(4.0 \times 10^{-6})^2$ \\ 
    \hline \hline
    \textbf{Coherence times} & $T_1$ & $T_2^{^*}$ & $T_2^{\mathrm{\,CPMG}}$ & $T_2^{\mathrm{\,CPMG}}$ / $T_2^{^*}$ \\
    \hline
    $\Phi_{\mathrm{b}} = 0   \; \textrm{m} \Phi_0$ & $12 \; \mu \textrm{s}$ & $2.5 \; \mu \textrm{s}$ & $23 \; \mu \textrm{s}$ ($N=1$) &   $9$ \\
    $\Phi_{\mathrm{b}} = 0.4 \; \textrm{m} \Phi_0$ & $12 \; \mu \textrm{s}$ & $0.27 \; \mu \textrm{s}$ & $13 \; \mu \textrm{s}$ ($N=200$) &  $48$ \\
    \hline
 \end{tabular}
 \end{center}
    \begin{quote}
    In Ramsey-fringe and Hahn-echo simulations, we describe the Gaussian noise distributions by their standard deviations, $\sigma_\lambda$, obtained by integrating the $1/f$ noises over the bandwidth given by the experimental protocol (cut-off frequencies $\omega^{\,\lambda}_\mathrm{ir}$ and $\omega^{\,\varepsilon}_\mathrm{uv}$, see text).
    At $\varepsilon=0$, the dephasing improvement under a Hahn echo is greater than the theory would suggest for $\delta\Delta$ $1/f$ noise that extends to high frequencies; the lower $\omega^{\,\Delta}_\mathrm{uv}$ gives consistency (see Appendix).
    The equivalent flux and normalized critical-current noise amplitudes, $A_{\lambda}$, are values derived from the Ramsey and echo data assuming a power law $1/f^{\alpha}$ with $\alpha=1$ and that all noise in $\varepsilon$ and $\Delta$ is flux and critical-current noise, respectively; they are consistent with previously reported values\cite{Yoshihara-PRL-06,vanHarlingen-PRB-04}.
    Using these parameters in simulations yielded agreement with $N$-pulse dynamical-decoupling data, consistent with equation~(\ref{eq:chi}).
    The coherence times are given at two bias points dominated by $\delta\Delta$ and $\delta\varepsilon$ noise, respectively.
    \end{quote}
\end{table}


We now turn to transverse noise, \emph{i.e.\@} $\delta \varepsilon$-noise at $\varepsilon=0$, and $\delta \Delta$-noise at $\varepsilon\gg\Delta$ (Fig.~\ref{fig:psd}, inset), at the qubit frequency $\omega_{01}$ responsible for energy relaxation $\Gamma_1$.
In the low-temperature limit, $k_\mathrm{B}T\ll\hbar\omega_{01}$, where the environment cannot excite the qubit, the golden-rule expression for $\Gamma_1$ in a weakly damped quantum two-level system is
\begin{equation} \label{eq:Gamma1}
 \Gamma_1 = \frac{\pi}{2} \sum_{\lambda \in \delta \varepsilon,\delta \Delta}
    \left( \frac{\partial\omega_{\perp'}}{\partial\lambda} \right)^2 S_\lambda(\omega_{01}) = \frac{\pi}{2} S(\omega_{01}),
\end{equation}
with $\partial\omega_{\perp'}/ \partial\lambda$ the qubit's sensitivity to transverse noise and $S(\omega_{01})$ the total PSD (see Appendix).

We apply a long ($\gg T_1$) microwave pulse to saturate the transition and monitor the energy decay to the ground state, using equation~(\ref{eq:Gamma1}) to determine $S(\omega_{01})$ over the frequency range $\Delta \leq \omega_{01} \leq 2\pi\times21$~GHz by tuning $\Phi_{\mathrm{b}}$.
At $\Phi_\mathrm{b}=0$ ($\omega_{01} = \Delta$) and using a measurement-repetition period $t_\mathrm{rep}>1\,$ms, we observe $T_1=12 \pm 1\,\mu$s as shown in Fig~\ref{fig:time_domain}d.
As $t_\mathrm{rep}$ becomes shorter than 1 ms, the decay becomes increasingly non-exponential, which we attribute to the residual presence of non-equilibrium quasiparticles generated by the switching SQUID during readout.
We observe structure in the $\Gamma_1$ data due to environmental modes (\emph{e.g.\@} cavity modes, impedance resonances) with uncontrolled couplings to the qubit (Fig.~\ref{fig:psd}).
For comparison, we plot the expected Johnson-Nyquist flux noise in Fig.~\ref{fig:psd} due to the $R = 1/G = 50\,\Omega$ environment mutually coupled with strength $M=0.02$~pH to the qubit via the microwave line,
\begin{equation} 
  S^\mathrm{JN}_\varepsilon(\omega) =  \frac{1}{2\pi} \left(\frac{\partial\varepsilon}{\partial\Phi}\right)^2
  M^2  \frac{2 \hbar \omega \, G}{1 - e^{-\hbar \omega / k_{\mathrm{B}} T}}.
\end{equation}
This known noise source falls about $100$ times below the measured PSD at $f_{01}=\Delta/2\pi$ (red dot in Fig.~\ref{fig:psd}) where the relaxation is due solely to $\delta \varepsilon$ noise, indicating that the dominant source of energy relaxation lies elsewhere.
The crossover $f_\mathrm{c}$ between the effective $1/f$- and $f$-type flux noises (Fig.~\ref{fig:psd}) occurs between $\Delta/2\pi$ and $k_\mathrm{B}T/h$, where $T=50$\,mK is the approximate electronic temperature of our device\cite{Astafiev-PRL-04,Shnirman-PRL-05}.

%
The dynamical-decoupling protocols demonstrated in this work comprise the same types of simple pulses that are used for quantum gate operations and therefore require little additional overhead to implement.
Integrating refocusing pulses into qubit control sequences, {\em e.g.\em}, by forming composite gates that incorporate both quantum operations and refocusing pulses, will lead to lower net error rates in systems limited by dephasing\cite{Kerman-PRL-08}.
Despite observing levels of $1/f$ flux and critical-current noise similar to those observed ubiquitously in superconducting qubits and SQUIDs\cite{vanHarlingen-PRB-04,Clarke-Nature-08},
we could mitigate this noise dynamically to increase the pure dephasing times beyond 0.1\,ms, more than a factor $10^4$ longer than the intrinsic pulse length.
However, despite having a remarkably long energy relaxation time $T_1 = 12 \; \mu$s, the transverse relaxation $T_2 \approx 2T_1$ was ultimately limited by it.
Dynamical decoupling protocols go a long way to refocusing existing levels of $1/f$ noise and achieve long coherence times, and the main emphasis is now on identifying and mitigating the noise source(s) that cause energy relaxation.
We note for further study that the PSD power law obtained experimentally by the CPMG technique, when extended to higher frequencies, falls within a factor two of the measured transverse noise, suggesting the possibility that the microscopic mechanism responsible for low-frequency dephasing may also play a role in high-frequency relaxation.


\section*{Appendix}

\makeatletter \renewcommand{\thefigure}{A\@arabic\c@figure} \renewcommand{\thetable}{S\@arabic\c@table} \makeatother

\setcounter{figure}{0} \setcounter{table}{0}

\subsection*{Measurement set-up.}
We performed our experiments at MIT, in a dilution refrigerator with a base temperature of 12\,mK.
The device was magnetically shielded with 4 Cryoperm-10 cylinders and a superconducting enclosure. All electrical leads were attenuated and/or filtered to minimize noise.

We used the Agilent E8267D microwave source, and employed the Tektronix AWG 5014 arbitrary waveform generator to create I/Q modulated pulses, and to shape the read-out pulse.

\subsection*{Description of the qubit.}
\begin{figure}[b!]
\begin{center}
\includegraphics[width=11cm]{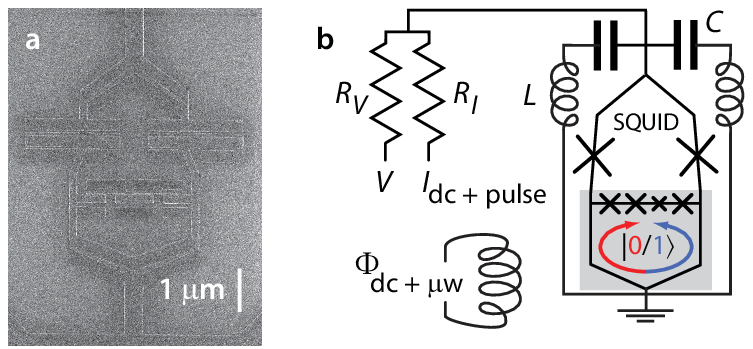}
\vspace{-5mm}
\begin{quote}
\caption{\textbf{a}, Scanning electron micrograph (SEM) of a device (qubit and SQUID shown) with identical design parameters as the one measured during this work~\cite{Yoshihara-PRL-06}.
    \textbf{b}, Schematic. Qubit loop (shaded) and galvanically coupled read-out SQUID.
    The crosses are Josephson junctions; $R$, bias resistors; $C$, shunt capacitances; $L$, inductances.
    }    \label{fig:device_schematic}
\end{quote}
\end{center}
\end{figure}

We fabricated our device at NEC, using the standard Dolan angle-evaporation deposition process of Al--AlO$_x$--Al on a SiO$_2$/Si wafer (Fig.~\ref{fig:device_schematic}a).

The persistent-current, or flux qubit \cite{Mooij-Science-99,Orlando-PRB-99,vdWal-Science-00,Chiorescu-Science-03} consists of a superconducting loop with diameter $d\sim2\,\mu$m, interrupted by four Josephson junctions (Fig.~\ref{fig:device_schematic}).
Three of the junctions each have the Josephson energy $E_\mathrm{J}=210$\,GHz, and charging energy $E_C=4$\,GHz; the forth is smaller by a factor $\alpha=0.54$.
The ratio of energy scales puts the device in the flux limit, $E_\mathrm{J}/E_C\approx50$, thus making the phases across the Josephson junctions well defined.
The geometric and kinetic loop inductances are negligible compared to the Josephson inductance:
$$
 \qquad \qquad  L_\mathrm{g}\sim\mu_0 d \sim 2\,\mathrm{pH}, \quad L_\mathrm{k} = \mu_0\lambda_\mathrm{L}^2 \, l/S \sim 30\,\mathrm{pH}, \quad L_\mathrm{J} = \Phi_0/2\pi I_\mathrm{c}\sim10\,\mathrm{nH},
$$
where we used $\lambda_\mathrm{L} = 100$\,nm,
$l = 10\,\mu$m, and
$S = 20 \times 250\,\mathrm{nm}^2$.

\begin{figure}[h!]
\begin{center}
\includegraphics[width=12cm]{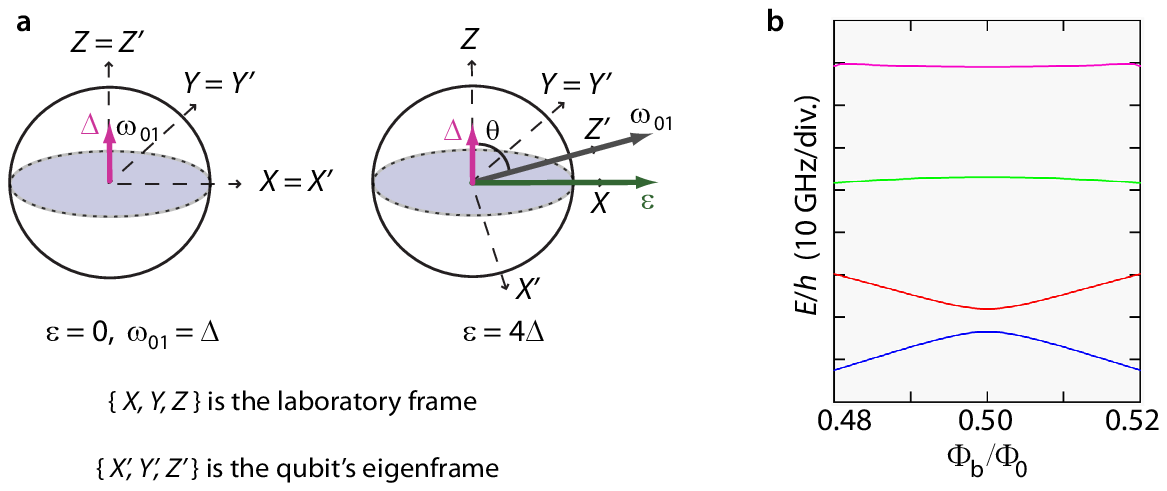}
\vspace{-5mm}
\begin{quote}
\caption{\textbf{Two-level Hamiltonian in different eigenbases, and simulated energy spectrum.}\newline
    \textbf{a}, The Hamiltonian (\ref{eq:Hamiltonian_labframe}) in the ``laboratory" frame $\{X,Y,Z\}$ has the $\varepsilon$ field (flux-bias) along $X$, perpendicular to the plane of the qubit loop, and the fixed tunnel coupling $\Delta$ along $Z$.
    The Hamiltonian (\ref{eq:Hamiltonian_eigenframe}) in the qubit's eigenframe $\{X',Y',Z'\}$ makes an angle $\theta$ with $Z$.
    The figure shows both frames under two bias conditions ($\varepsilon=0$ [$\theta=0^\circ$] and $\varepsilon=4\Delta$), with the laboratory frame fixed in space.
    The two frames coincide when $\varepsilon=0$.
    \textbf{b}, Simulated energy spectrum.
    }    \label{fig:lab_and_qubit_frames}
\end{quote}
\end{center}
\end{figure}

When the external magnetic flux $\Phi$ threading the qubit loop is close to half a magnetic-flux quantum, $\Phi_0/2$, the qubit's potential energy exhibits a double-well profile with quantized energy levels $\pm\hbar\varepsilon/2=\pm I_\mathrm{p}\Phi_\mathrm{b}$, where $\Phi_\mathrm{b}=\Phi-\Phi_0/2$ is the flux bias and $I_\mathrm{p}=0.18\,\mu$A the persistent current.
The ground states in the left and right potential wells correspond to the diabatic states $|L\rangle$ and $|R\rangle$
of opposite circulating persistent currents.
A tunnel coupling $\hbar\Delta/2$ between the right- and left-well qubit states opens an energy gap, $\hbar\Delta$, between the ground and excited states, $|0\rangle$ and $|1\rangle$, at flux degeneracy, $\Phi_\mathrm{b}=0$.
The two-level Hamiltonian -- analogous to a spin-1/2 particle in a magnetic field -- is in the ``laboratory" frame
\begin{equation} \label{eq:Hamiltonian_labframe}
    \hat{\mathcal{H}} = - \frac{\hbar}{2} \left[ (\varepsilon + \delta \varepsilon) \hat{\sigma}_{x} + (\Delta + \delta \Delta) \hat{\sigma}_{z} \right],
\end{equation}
where $\hbar\omega_{01} = \hbar\sqrt{\varepsilon^2+\Delta^2}$ is the energy-level splitting; $\hat{\sigma}_j$ are the Pauli matrices; and $\delta \varepsilon$ and $\delta \Delta$ are the noise fluctuations.
The quantization axis makes an angle $\theta=\arctan(\varepsilon/\Delta)$ with $\hat{\sigma}_{z}$ (and an angle $\pi-\theta$ with the persistent-current eigenstates), so that the qubit is first-order insensitive to flux noise when biased at the ``sweet spot" $\varepsilon=0$.
(We write the Hamiltonian in this way rather than with swapped $x$ and $z$ indices as in several previous papers, so that our pulses will be along $X$ and $Y$ in agreement with the habitual NMR language.)
It can, alternatively, be written in the qubit's eigenbasis,
\begin{equation} \label{eq:Hamiltonian_eigenframe}
 \hat{\mathcal{H}} = -\frac{1}{2} \hbar \left( \omega_{01}\hat{\sigma}_{z'} + \delta\omega_{z'} \hat{\sigma}_{z'} + \delta\omega_{\perp'} \hat{\sigma}_{\perp'} \right) ,
\end{equation}
see Fig.~\ref{fig:lab_and_qubit_frames}a.
Here $\hat{\sigma}_{\perp'}$ denotes that the transverse spin component can include both $\hat{\sigma}_{x'}$ and $\hat{\sigma}_{y'}$.
The longitudinal and transverse noises,
$\delta\omega_{z'}$ and $\delta\omega_{\perp'}$, are further described below.

\clearpage

\subsection*{Description of the read-out SQUID.}

Our hysteretic dc-SQUID \cite{SQUID-handbook}, see Fig.~\ref{fig:device_schematic}, has critical current $I_\mathrm{c}=4.5\,\mu$A; normal resistance $R_N = 0.25\,\mathrm{k}\Omega$; mutual qubit--SQUID inductance $M_{\mathrm{Q-S}}=21$\,pH; on-chip capacitors $C\sim10$\,pF, inductors $L\sim0.1$\,nH, and bias resistors $R_I=0.2\,\mathrm{k}\Omega$ and $R_V=1\,\mathrm{k}\Omega$;
and further cold and room-temperature resistors and filters.
A coil with mutual coupling $M_\mathrm{dc}=0.6$\,pH to the qubit provides the dc flux bias, and an on-chip antenna with an estimated $M_{\mu\mathrm{w}}\approx0.1$\,pH the microwave excitation.
The shunt capacitors bring the plasma frequency down to $\omega_\mathrm{p}/2\pi=2.1$\,GHz.

\begin{figure}[b!]
\begin{center}
\includegraphics[width=10cm]{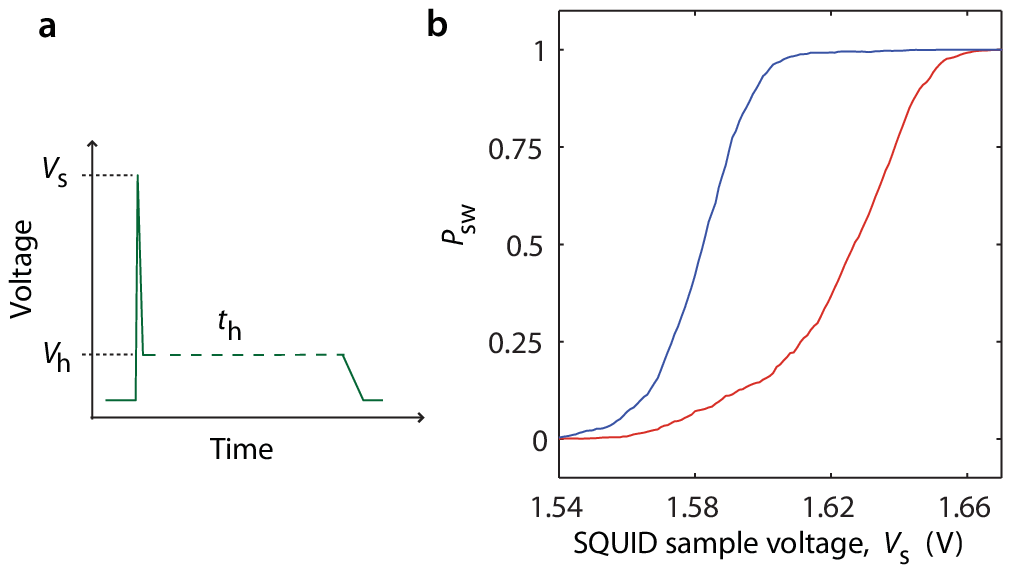}
\vspace{-5mm}
\begin{quote}
\caption{\textbf{a}, Read-out pulse (voltage across a 1-k$\Omega$ bias resistance). The optimized pulse shape is programmed to have a 4-ns sample pulse, including 1-ns rise, 1-ns sample, and 2-ns fall times, followed by a 3-$\mu$s hold plateau at 17.5\% of the sample voltage.
    \textbf{b}, Read-out visibility  at $\Phi_\mathrm{b}=0$.
    Scans of the SQUID's switching probability, $P_\mathrm{sw}$, vs.\@ the height of the sampling pulse.
    We obtain 79\,\% read-out visibility between the qubit's ground and excited states.
    Relaxation during read out leads to an imperfect determination of the excited state (17\,\% dark counts at the optimal $V_\mathrm{s}$).
    }    \label{fig:ReadOut_Scurves}
\end{quote}
\end{center}
\end{figure}

The qubit's magnetization, resulting from the persistent currents, modulates the SQUID's switching current.
For qubit read out, we apply a sample-and-hold current pulse to the SQUID (Fig.~\ref{fig:ReadOut_Scurves}a), and use a threshold detector, after room-temperature amplification, to register the presence or absence of a voltage, conditioned on the qubit being in state $|L\rangle$.
We determine the switching probability, $P_\mathrm{sw}$, statistically by repeating this measurement several thousand times.

The pulse is produced by a digital arbitrary-waveform generator (Tektronix AWG5014) with 250 MHz analog output bandwidth.
Due to enhanced relaxation when a current is flowing through the SQUID, a rapid sample pulse is important for good read-out visibility.
A hold current enables the room-temperature electronics to register the voltage pulse before retrapping occurs, but a low and short hold pulse limits the quasiparticle generation, which is also beneficial for the visibility.
We obtain an optimal read-out visibility of 79\% (Fig.~\ref{fig:ReadOut_Scurves}b).

In addition to biasing the SQUID, the read-out pulse couples flux into the qubit with the effect of shifting the states adiabatically, $|0\rangle\to|L\rangle$ and $|1\rangle\to|R\rangle$, before the measurement takes place, so that the states can be distinguished by the projective measurement in the persistent-current basis.

The optimal SQUID dc-bias current $I_\mathrm{b}$ with respect to noise coupled into the qubit was very close to $I_\mathrm{b}=0$, indi\-ca\-ting highly symmetric SQUID junctions.
In this set of experiments, we therefore did not apply any dc bias to the SQUID while manipulating the qubit, \emph{cf.\@} Ref.~\cite{Yoshihara-PRL-06}.


\subsection*{Sensitivity to noise.}


Following the approach of Ithier \emph{et al.} \cite{Ithier-PRB-05}, we will evaluate the noise terms in the qubit's Hamiltonian,
$$
 \hat{\mathcal{H}} = -\frac{1}{2} \hbar \left( \omega_{01}\hat{\sigma}_{z'} + \delta\omega_{z'} \hat{\sigma}_{z'} + \delta\omega_{\perp'} \hat{\sigma}_{\perp'} \right) \qquad \qquad \qquad \mathrm{[equation~(\ref{eq:Hamiltonian_eigenframe})]}.
$$
We consider both flux and critical-current noise (affecting $\Delta$),
and express the noises as sensitivity derivatives that translate the noise fluctuations to a change in the Hamiltonian, \emph{i.e.\@} $\delta\lambda\to\delta\omega$.
At $\Phi_{\mathrm{b}}=0$, the first-order noise $\delta \Delta$ is much larger than second-order $\delta \varepsilon$ (flux) noise, as confirmed through simulation.
It is therefore sufficient to expand to first order to explain our Ramsey and echo data,
\begin{equation} \label{eq:long_noise_comp}
 \delta\omega_{z'} = \frac{\partial\omega_{01}}{\partial\lambda} \delta\lambda + \ldots
 \qquad \mathrm{and} \qquad
  \delta\omega_{\perp'} = \frac{\partial\omega_{\perp'}}{\partial\lambda} \delta\lambda + \ldots \, .
\end{equation}
%

Energy relaxation is related to noise that is transverse to the qubit's quantization axis, $\delta\omega_{\perp'}(\lambda)$, at the frequency of the level splitting $\omega_{01}$.

Pure dephasing, on the other hand, is related to low-frequency fluctuations of the qubit's energy-level splitting.
%
To evaluate the longitudinal first-order term in equation~(\ref{eq:long_noise_comp}) we use the chain rule,
\begin{equation}  
 \frac{\partial\omega_{01}}{\partial\lambda} \, = \, \frac{\partial\omega_{01}}{\partial\varepsilon} \frac{\partial\varepsilon}{\partial\lambda} \, + \, \frac{\partial\omega_{01}}{\partial\Delta} \frac{\partial\Delta}{\partial\lambda} .
\end{equation}
By geometry,
\begin{equation}
 \frac{\partial\omega_{01}}{\partial\varepsilon} = \frac{\varepsilon}{\omega_{01}} \qquad \mathrm{and} \qquad \frac{\partial\omega_{01}}{\partial\Delta} = \frac{\Delta}{\omega_{01}} .
\end{equation}
From spectroscopy measurements we infer the $\varepsilon$ sensitivity to flux noise, $\lambda=\Phi$,
\begin{equation}
  \xi = \frac{\partial\varepsilon}{\partial\Phi} = 2\pi\times 1.1\,\mathrm{GHz/m}\Phi_0 ,
\end{equation}
while $\Delta$ is insensitive to $\Phi$ noise.

Since $\hbar\varepsilon=2 \,I_\mathrm{p} \, \Phi_\mathrm{b}$ and
$I_\mathrm{p}=I_\mathrm{c}\sqrt{1-1/(2\alpha)^2}$,\, $\varepsilon$ is sensitive also to critical-current noise, $\lambda=I_\mathrm{c}$,
\begin{equation}
  \frac{\partial\varepsilon}{\partial I_\mathrm{c}} = \frac{\varepsilon}{I_\mathrm{c}} .
\end{equation}
For the $\Delta$ sensitivity to fluctuations in $i_\mathrm{c}=\delta I_\mathrm{c}/I_\mathrm{c}$,
a numerical simulation gives $\partial\Delta/\partial I_\mathrm{c} = 2\pi \times 8\cdot10^{15}$\,Hz/A,
and with $I_\mathrm{c}=0.4\,\mu$A we obtain
\begin{equation}
  \kappa_1 = \frac{\partial\Delta}{\partial i_\mathrm{c}} = 2\pi \times 3.01\,\mathrm{GHz} .
\end{equation}
Taken together, the first-order fluctuations are therefore
\begin{equation}
 \frac{\partial\omega_{01}}{\partial\lambda} \, \delta\lambda \, = \, \frac{\varepsilon}{\omega_{01}}\xi \, \delta\Phi_\mathrm{b} \, + \, \frac{\varepsilon}{\omega_{01}}\frac{\varepsilon}{I_\mathrm{c}} \, \delta I_\mathrm{c} \, + \, \frac{\Delta}{\omega_{01}}\kappa_1 .
\end{equation}
Flux noise dominates, except very near $\epsilon=0$, where the $\delta\Delta$ noise is needed to account for the observed decay. (The second term is negligible.)

\clearpage

\subsection*{Decoherence. Ramsey and echo phase decays --- $1/f$-flux noise.}

In the Bloch--Redfield formalism, valid for weakly coupled, short-correlated noise, the dynamics of two-level systems is described by the longitudinal and transverse relaxation rates, $\Gamma_1 = 1/T_1\,$ and $\Gamma_2 = 1/T_2\,$, respectively~\cite{Geva-Skinner-JChemPhys-95,Ithier-PRB-05}.
We assume that the qubit is coupled to many fluctuators, which, in concert and independent of their individual statistics, will yield a Gaussian noise distribution due to the central limit theorem.
In systems where the noise is regular at the frequency of the qubit's energy-level splitting, $\omega_{01}$, relaxation and dephasing factorize.
The longitudinal relaxation is exponential due to the many uncorrelated contributions of transversally coupled noise at $\omega_{01}$.
The pure-dephasing rate, $\Gamma_\varphi$, associated with low-frequency, quasi-static, longitudinally coupled noise (inhomogeneous broadening), combines with $\Gamma_1$ to give the rate $\Gamma_2 = \Gamma_1/2 + \Gamma_\varphi$.
However, the exact form of the time dependence of the dephasing component is determined by the noise-PSD.
(That is, when the dephasing is non-exponential, the inverse time constant can strictly no longer be interpreted as a rate.)

\begin{figure}[b!]
\begin{center}
\includegraphics[width=12cm]{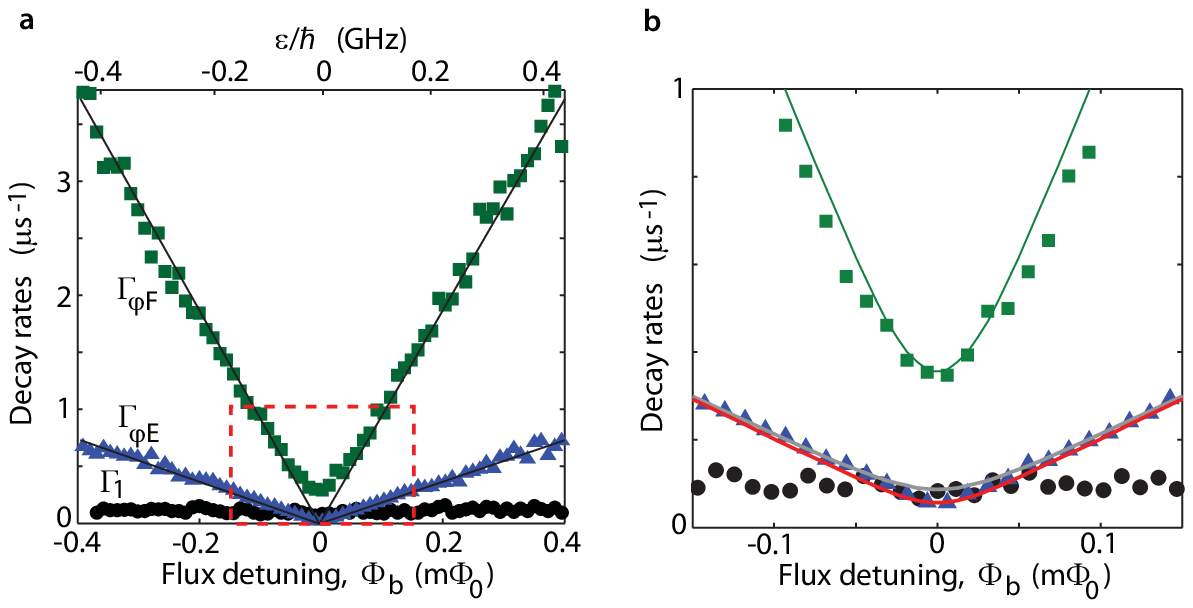}
\vspace{-5mm}
\begin{quote}
\caption{\textbf{Decay rates vs.\@ flux detuning.  a},
    Energy-relaxation rate, $\Gamma_1$, and Gaussian phase-decay rates for the Hahn echo, $\Gamma_{\varphi,\mathrm{E}}$, and Ramsey free induction, $\Gamma_{\varphi,\mathrm{F}}$, after subtracting the exponential $\Gamma_1$ decay.
    Straight, black lines are fits to $\Gamma_{\varphi, \mathrm{E/F}}(\Phi_\mathrm{b})$ for flux noise only; see the text and equation~(\ref{eq:phase_decays}).
    \textbf{b}, Blow up of the data in the dashed-box region in \textbf{a}, along with simulated Ramsey- and echo- decay rates including both $\delta\varepsilon$ and $\delta\Delta$ noises.
    For the echo, the grey line is the rate obtained for a $1/f$ noise in $\delta\Delta$ with the 1-MHz ultra-violet cut off given by the experimental protocol.
    The red line is the rate obtained with the lower ultra-violet cut off: $\omega_\mathrm{uv}^\lambda/2\pi < 0.1\,$MHz. The exact rate depends on the detailed shape of the noise cut off, which is unknown.
    }    \label{fig:Decays_vs_detuning}
\end{quote}
\end{center}
\end{figure}

At $\Phi_\mathrm{b}=0$, the echo decay is nearly $T_1$ limited in our device, and therefore practically indistinguishable from an exponential. Relaxation contributes by $\Gamma_1=1/2\,T_1=43\times10^3$/s or 6.8\,kHz to the low-power spectroscopic line width $\Delta f_\mathrm{(FWHM)}=0.18$\,MHz.
This exceeds the expected
$\Delta f_\mathrm{(FWHM)}=1/\pi\,T_2^{^*}$ by only 0.05\,MHz, indi\-ca\-ting little power broadening, given the free-induction decay rate, $\Gamma_\mathrm{F}=1/T_2^{^*}$, with the measured $T_2^{^*}=2.5\,\mu$s.

Biased away from $\Phi_\mathrm{b}$, we find a Gaussian spin-echo phase decay, $\exp[-(\Gamma_{\varphi,\mathrm{E}}\,t)^2]$, consistent with a $1/f$ flux-noise model \cite{Ithier-PRB-05} (singular PSD at $\omega=0$).
As the Ramsey fringe decays considerably faster it is hard to distinguish between exponential and Gaussian decays.
The bias dependencies for echo and Ramsey decays are
\begin{equation} \label{eq:phase_decays}
  \Gamma_{\varphi,\mathrm{E/F}}(\Phi_\mathrm{b}) = (A_\Phi\,\eta_\mathrm{\,E/F})^{1/2} \, \left| \frac{\partial\omega_{01}}{\partial\Phi} \right| ,
\end{equation}
where the numerical factors $\eta$ differ due to the different echo and Ramsey filtering functions: $\eta_{\varphi,\mathrm{F}} = \ln(1/\omega_\mathrm{ir}t)$
and $\eta_{\varphi,\mathrm{E}} = \ln 2$.
Their ratio,
$\Gamma_{\varphi,\mathrm{F}}(\Phi_\mathrm{b}) / \Gamma_{\varphi,\mathrm{E}}(\Phi_\mathrm{b}) \approx 4.5$,
is in accordance with our data, and the magnitude,
\begin{equation} \label{eq:A_phi}
 A_\Phi=(1.7\,\mu\Phi_0)^2 ,
\end{equation}
of the noise $S(\omega)=A_\Phi/|\omega|$ agrees with previous results \cite{Yoshihara-PRL-06}, see Fig.~\ref{fig:Decays_vs_detuning}.


\subsection*{Numerical simulations of Ramsey and echo.}

The $\Phi_\mathrm{b}$ dependencies of $T_2^{^*}$ and $T_{2,\mathrm{E}}$ are reproduced in numerical simulations.
We simulated the Ramsey fringe by numerically solving the Bloch equations, including the measured $T_1$ decay and linearly coupled quasi-static noises,
averaging over many realizations.
Each run had stochastic, normally distributed $\delta\varepsilon$ and $\delta\Delta$ deviations from their average values:
The $1/f$ flux noise, $S(\omega)=A_\Phi/|\omega|$, gives
$\sigma_\varepsilon/2\pi = 10$\,MHz when integrated from 1~Hz to 1~MHz;
noise in $\Delta$ dominates near $\Phi_\mathrm{b}=0$, and we get agreement for $\sigma_\Delta/2\pi = 0.06$\,MHz, uncorrelated with the flux noise and obtained from
$S_\Delta(\omega) = A_\Delta/|\omega|$ with
$A_\Delta = \left(4\cdot10^{-6}\right)^2$.

In the echo simulation we additionally took into account dynamic noises from $10^4$ to $10^{10}$ Hz.
We obtained the noisy time series for $\varepsilon$ and $\Delta$ by inverse-Fourier transforming the amplitudes of the $1/f$ noises with random phases for each Fourier component,
and then evaluated the Schr\"{o}dinger evolution operator in discrete time.
%

Near $\Phi_\mathrm{b}=0$, the experimental data shows a greater echo-improvement in $T_2$ than the simulation would suggest for these parameters (see Fig.~\ref{fig:Decays_vs_detuning}b).
This discrepancy would be explained by a lower ultra-violet $\Delta$-noise cut off, with faster decay than $1/f$ above $\omega_\mathrm{uv}^{\Delta}/2\pi \sim 0.1$\,MHz.

Numerical evaluations of the coherence integral, equation~(\ref{eq:chi_1}) below, agree with our simulations.


\subsection*{Pulse calibration.}

We calibrate the rotations to $<1\,\%$ accuracy by applying a tune-up sequence of pulses, akin to methods used in NMR~\cite{Vaughan-RSI-72}.
A rigorous measurement of gate errors should be done with randomized benchmarking, for example.

\subsection*{Dynamical-decoupling pulse sequences.}

Collin \emph{et al.} \cite{Collin-PRL-04} and Ithier \emph{et al.} \cite{Ithier-PRB-05} have employed some NMR methods beyond the Hahn spin echo to manipulate a superconducting qubit.
In this work, we apply multi-pulse, dynamical-decoupling pulse sequences to significantly enhance the coherence times and to facilitate spectroscopy of the environmental noise.

\begin{figure}[h!]
\begin{center}
\includegraphics[width=13cm]{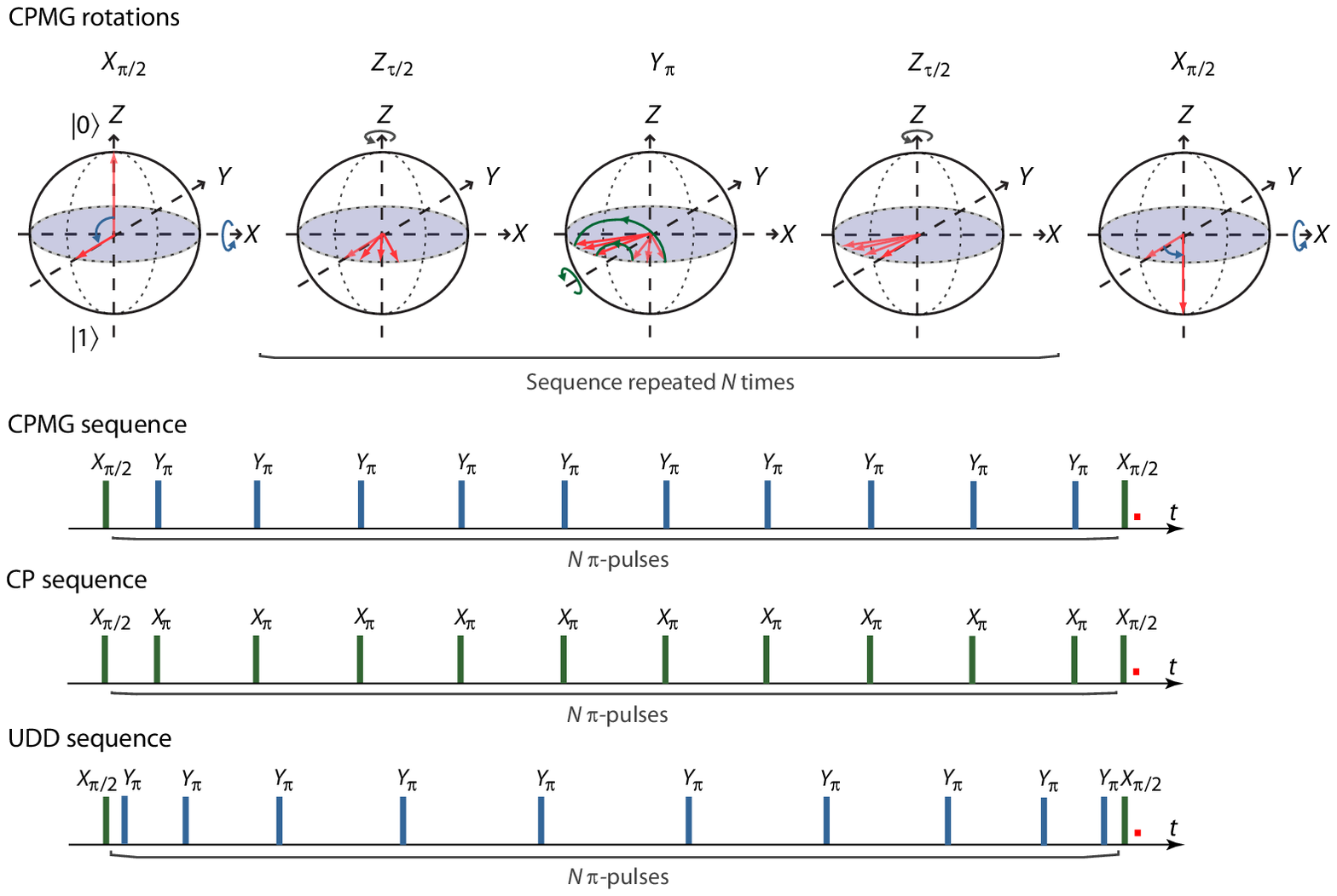}
\vspace{-5mm}
\begin{quote}
\caption{\textbf{Dynamical-decoupling pulse sequences.} Rotations of the Bloch vector during the CPMG sequence (pulses along $X$ and $Y$ in the laboratory frame.)
    Timing of the CPMG, CP, and UDD pulse sequences for $N=10$.
    }    \label{fig:CPMG_rotations}
\end{quote}
\end{center}
\end{figure}

During the Carr-Purcell-Meiboom-Gill (CPMG) multi-echo sequence \cite{Carr-Purcell-PR-54,Meiboom-Gill-RSI-58}, defined as
\begin{equation} \label{eq:CPMG_sequence}
 X_{\pi/2} - \Big( \frac{\tau}{2N} - Y_\pi - \frac{\tau}{2N} \Big)_N - X_{\pi/2}
\end{equation}
and illustrated by the rotations in the Bloch sphere (Fig.~\ref{fig:CPMG_rotations}), the transverse component of the Bloch vector is refocused along the axis of the refocusing pulses (here $Y$), whereas the perpendicular component ($X$) is randomized.
CPMG is a development on the Carr-Purcell (CP) sequence \cite{Carr-Purcell-PR-54}, which has identical pulse positions, but where all rotations are along the same axis.
The third pulse sequence that we investigated, the Uhrig dynamical-decoupling (UDD) sequence \cite{Uhrig-PRL-07,Uhrig-NJPhys-08}, has normalized pulse positions defined by
\begin{equation} \label{eq:UDD_sequence}
 \delta_j = \sin^2 \left( \frac{\pi j}{2N+2} \right).
\end{equation}

The dephasing under a certain sequence is described by the \emph{coherence integral},
\begin{equation} \label{eq:chi_1}
  \chi_N(\tau) \, = \, \left(\frac{\partial\omega_{01}}{\partial\lambda}\right)^2 \tau^2 \,
  \int_0^\infty \mathrm{d}\omega \, S(\omega) \,
  g_N(\omega,\tau) .
\end{equation}
The dimensionless
filter function $g_N(\omega,\tau)$ depends on the number and distribution of $\pi$ pulses \cite{Uhrig-PRL-07,Uhrig-NJPhys-08,Cywinski-PRB-08,Biercuk-Nature-09,Biercuk-PRA-09},
\begin{equation} \label{eq:filter_def}
  g_N(\omega,\tau) = \frac{|y_N(\omega,\tau)|^2}{(\omega\tau)^2},
\end{equation}
\begin{equation} \label{eq:filter_def_2}
  y_N(\omega,\tau) = \left| 1+(-1)^{1+N}\exp(i\omega\tau) + 2\sum_{j=1}^{N} (-1)^j \exp(i\omega\delta_j\tau) \cos(\omega\tau_\pi/2) \right| ,
\end{equation}
where $\delta_j \in [0,1]$ is the normalized position of the centre of the $j$th $\pi$ pulse between the two $\pi/2$ pulses.
The $\tau_\pi$-dependent factor assumes square pulses, but is a reasonable approximation in our case.
Alternatively to equation~(\ref{eq:filter_def}), one can define a function $F_N(\omega,\tau) = |y_N(\omega,\tau)|^2$ that filters the phase noise, $S(\omega)/\omega^2$, as in, \emph{e.g.\@}, Refs.~\cite{Uhrig-PRL-07,Uhrig-NJPhys-08,Cywinski-PRB-08,Biercuk-Nature-09,Biercuk-PRA-09}.

The Ramsey free induction ($N=0$) and the Hahn spin echo ($N=1$) have weighting functions
$g_0(\omega,\tau) = \mathrm{sinc}^2(\omega\tau/2)$ and $g_1(\omega,\tau) = \mathrm{sinc}^2(\omega \tau/4) \sin^2(\omega \tau/4)$, respectively.
Note that $g_1(0)=0$, indicating a suppression of the low-frequency part of the noise for the Hahn echo.

The CPMG sequence is inherently robust to field inhomogeneities when the following criteria are fulfilled \cite{Borneman-JMR-10}: %
(i) The effective rotation axis is oriented in the transverse ($XY$) plane (true when the driving frequency is resonant with the level splitting);
(ii) the magnetization (Bloch vector) is initially aligned with the rotation axis ($Y$);
and (iii) the rotation angle is $\pi$.
Errors that occur due to deviations from (i--iii) can be quantified by computing the propagator over one cycle of the sequence.
For an initial state $Y$, one finds that $Y_\pi$ errors appear only to fourth order (CPMG), whereas $X_\pi$ errors accumulate to second order (CP).

\subsection*{Determination of the PSD from the CPMG-decay data.}

The qubit is subjected to different decay mechanisms during the CPMG pulse sequence.
We identify three decay rates, and fit the qubit's population to the function
\begin{equation} \label{eq:P_FitFcn}
  P_\mathrm{sw}(\tau) = P_0 + a
  \exp\left( -\Gamma_1 \tau \right) \,
  \exp\left( -\Gamma_\mathrm{p} (\tau_\mathrm{p}) \right) \,
  \exp\left( -\chi_N(\tau) \right) ,
\end{equation}
keeping the constants $a$ and $P_0$ fixed for all $N$.
%
\begin{itemize}
\item {\bf Relaxation. \bf}
For the duration of the sequence, there is energy relaxation with the constant rate $\Gamma_1 = 1/2\,T_1$, where $T_1=12\,\mu$s.
\item {\bf Decay during pulses. \bf}
In addition, there is pulse-induced decoherence during the total pulse time $\tau_\mathrm{p} = N\tau_\pi+2\,\tau_{3\pi/2}$.
(For experimental reasons, instead of $\pi/2$ pulses we often use the equivalent $3\pi/2$ pulses.)
We do not need to assume that this decay has any particular form, \emph{e.g.} exponential, as it will be divided out of the calculation.
However, a simple exponential decay with $\Gamma_\mathrm{p}=1/(1.75\,\mu$s) gives a good overall fit independent of $N$.
Although much higher than the Rabi-decay rates described in the manuscript, this rate is reasonable given that we observe an increased Rabi-decay time at high driving amplitudes.
\item {\bf Dephasing. \bf}
The fit parameter $\Gamma_\varphi^{(N)}$ is the dephasing rate during the total free-evolution time $\tau$ under an $N$-pulse CPMG sequence, which provides us with a means of figuring the noise PSD, $S(\omega)$.
Only the PSD appears in the decay function, because the noise statistics were taken to be Gaussian, and all information is included in the second-order correlation function.
\end{itemize}


Dividing equation~(\ref{eq:P_FitFcn}) with itself for two free-evolution times, $\tau_1$ and $\tau_2$, we get
\begin{equation} \label{eq:P_FitFcn_Ratio}
  q(\tau_1,\tau_2) = \frac{P_\mathrm{sw}(\tau_1) - P_0}{P_\mathrm{sw}(\tau_2) - P_0} =
  \exp\Big(-\Gamma_1 [\tau_1-\tau_2] \Big) \,
  \exp\Big(-[\chi_N(\tau_1)-\chi_N(\tau_2)] \Big) .
\end{equation}

\begin{figure}[h!]
\begin{center}
\includegraphics[width=13cm]{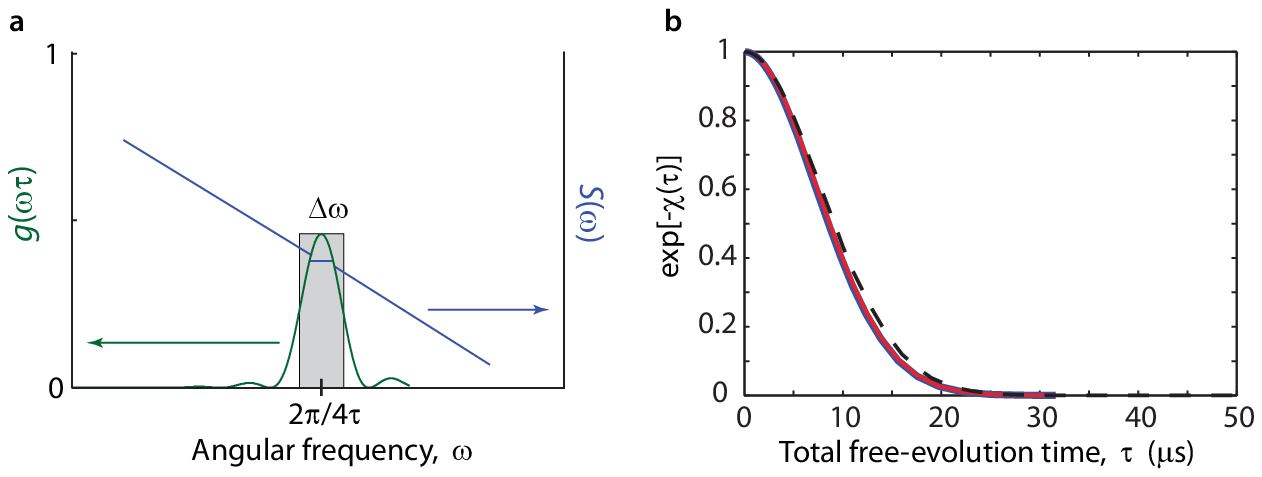}
\vspace{-5mm}
\begin{quote}
\caption{\textbf{a}, The filter function is narrow, and the noise PSD is smooth enough, that we can obtain the noise PSD at the filter's centre frequency by evaluating the coherence integral $\chi_N(\tau)$, equation~(\ref{eq:chi_2}).
   \textbf{b},
   Fitted phase decay, $\exp[-\chi_N(\tau)]$, under an $N\!=\!84$ $\pi$-pulse CPMG sequence with $\tau_\pi = 6.8$\,ns, at the flux detuning  $\varepsilon/2\pi=-430$\,MHz ($\Phi_0=-0.4\,\mathrm{m}\Phi_0$).
   \textbf{Blue} is the function $\exp(-(\Gamma_\varphi^{(N)} \tau)^2)$ with a fitted $\Gamma_\varphi^{(N)}=8.5\,\mu$s.
   \textbf{Red} is the integral (\ref{eq:chi_1}) with $S(\omega)=A/\omega$.
   Note the excellent agreement between the blue and red lines.
   \textbf{Dashed} is the expression (\ref{eq:chi_2}) with the same parameters as for the red line.
   }\label{fig:Filter_with_1_f}
\end{quote}
\end{center}
\vspace{-1cm}
\end{figure}

For a pulse separation $\tau' = \tau/N$ (and large $N$), the filter (\ref{eq:filter_def}) is peaked near
\begin{equation} \label{eq:filter_center_fq}
  \omega'(\tau') = \frac{2\pi}{4\tau'} .
\end{equation}
It is narrow enough, that we can treat $g_N\Big(\omega'(\tau'),\tau'\Big)$ as a delta function peaked at $\omega'$, and regard the noise in the qubit's transition frequency as constant within its bandwidth, $\Delta\omega$, as illustrated in Fig.~\ref{fig:Filter_with_1_f}a. This provides us with a means of characterizing the noise at a frequency $\omega$ by varying $\tau$.
We can rewrite the integral in equation~(\ref{eq:chi_1}) as
$\int_0^\infty \!\mathrm{d} \, \omega \, g_N\Big(\omega'(\tau') ,\tau'\Big) \to \Delta\omega \, \int_0^\infty \!\mathrm{d} \,\omega \, \delta(\omega-\omega')$,
and do the approximation
\begin{equation} \label{eq:chi_2}
  \chi_N(\tau) \approx \left(\frac{\partial\omega_{01}}{\partial\lambda}\right)^2 \tau^2 S\Big(\omega'(\tau')\Big) \, g_N\Big(\omega' (\tau') ,\tau'\Big)  \Delta\omega .
\end{equation}
We know $\Delta\omega(\tau)$, and $g_N\Big(\omega'(\tau'),\tau'\Big)$ from a numerical calculation for each $N$.

We can use equation~(\ref{eq:chi_2}) to evaluate the second factor on the right-hand side of equation~(\ref{eq:P_FitFcn_Ratio}),
\begin{equation} \label{eq:P_FitFcn_Ratio_chi}
  \tilde{q}(\tau_1, \tau_2) \, = \, \exp
  \left(-
  \left( \frac{\partial\omega_{01}}{\partial\lambda} \right)^2 \left[
  \tau_1^2 \, S\Big(\omega'(\tau_1)\Big) \, g_{N,1}\Big(\omega'(\tau_2),\tau_2\Big) \, \Delta\omega_1
  -
  \tau_2^2 \, S\Big(\omega'(\tau_2)\Big) \, g_{N}\Big(\omega'(\tau_2),\tau_2\Big) \, \Delta\omega_2
  \right]
  \right) .
\end{equation}
This gives us the difference between the noises at $\omega'(\tau_1)$ and $\omega'(\tau_2)$, so that we can find one if we know the other.
To get the absolute noise level, we have to assume $S(\omega)$ for some value of $\omega'=\omega'(\tau')$, \emph{e.g.}, a value consistent with the PSD found in the spin-echo measurements, equation~(\ref{eq:A_phi}), or indeed choose $\chi_N=1$ for $\tau=0$ and go from there.
Following this procedure, we obtain a number of points representing the noise PSD over 0.2--20 MHz.
The points within the range 0.2--2~MHz lie tight together and are the least susceptible to error, which is why we fit this data to a straight line, rendering the reported $1/f^\alpha$ dependence: $A'_\varepsilon = (0.8\,\mu\Phi_0)^2$ and $\alpha=0.9$.
Integrating the noise PSD from these parameters yields $\sigma_\varepsilon=8$~MHz (compared to 10\,MHz) for the quasi-static noise, slightly changing the slope of the calculated $\Gamma_{\varphi,\mathrm{F}}$ in Fig.~\ref{fig:Decays_vs_detuning}, which presumed $\alpha=1$.
However, the slope of the PSD is not necessarily constant over many orders of magnitude, which could explain the slight deviation for the Ramsey, being sensitive to low-frequency noise.

Alternatively, we can fix the rates $\Gamma_\mathrm{p}$, $\Gamma_1$, and then fit the remaining, Gaussian decay $\chi_N(\tau) = (\Gamma_\varphi\tau)^2$.
Then each measured point, $P_\mathrm{sw}(\tau)$, is mapped onto $S(\omega'(\tau))$ by using equation~(\ref{eq:chi_2}).
However, the using a Gaussian decay function is equivalent to an \emph{a priori} assumption of $\alpha=1$.

Figure~\ref{fig:Filter_with_1_f}b shows a numerical evaluation of the coherence integral (\ref{eq:chi_1}), assuming a noise spectral density
\begin{equation} \label{eq:1_f}
 S(\omega) = A/\omega .
\end{equation}
The calculation shows, that this method underestimates the decay by about 5\,\%, since it disregards the part of the noise that falls within the harmonics of $g_N$, the first of which occurs at $3\omega'$ and which is about 10\,dB lower than the main lobe at $\omega'$.

\subsection*{Decoherence during driven evolution.}

We obtain information about the flux noise in the 2--20\,MHz frequency range also by analyzing the decoherence during driven evolution.


Driven by a near-resonant, transverse field
$\hat{\mathcal{H}}_1 = -\hbar\,\Omega(t)\,\cos(\omega t)\,\hat{\sigma}_x$,
the static part of the Hamiltonian in the rotating frame is
\begin{equation} \label{eq:Hamiltonian_rot_frame}
 \hat{\tilde{\mathcal{H}}} = -\frac{1}{2} \hbar \left(
 \Delta\omega \hat{\sigma}_z + \Omega \hat{\sigma}_x \right) ,
\end{equation}
where $\Delta\omega = \omega - \omega_{01}$.
As a result, the qubit oscillates at the Rabi frequency
\begin{equation} \label{eq:Rabi_fq}
 \Omega_\mathrm{R} = \sqrt{\Omega^2 + (\Delta\omega)^2} \approx \Omega + (\Delta\omega)^2/2\Omega
\end{equation}
around the effective field, which makes an angle $\eta=\arctan(\Delta\omega/\Omega)$
with the qubit's $\hat{\sigma}_z$ axis.

The decay of the Rabi oscillations has contributions from noise at the frequencies $\omega_{01}$ and $\Omega$ \cite{Ithier-PRB-05}, and from the low-frequency, quasi-static flux noise with variance $\sigma_\varepsilon^2$.

\paragraph{Static noise.}
We first describe the decay due to quasi-static noise.
Flux deviations from the bias point $\varepsilon$ lead to a frequency detuning
$\Delta\omega = (\partial\omega_{01}/\partial\varepsilon) \Delta\varepsilon = (\varepsilon/\omega_{01})\Delta\varepsilon$,
normally distributed with a standard deviation $\sigma_\varepsilon$,
\begin{equation} \label{eq:Noise_distr_Gaussian}
N(\Delta\varepsilon) = \sigma_\varepsilon^{-1/2} \exp\left(-(\Delta\varepsilon)^2/2\,\sigma_\varepsilon^2 \right) .
\end{equation}
Integrating over $N(\Delta\varepsilon)$, the Rabi oscillations decay as
\begin{equation} \label{eq:Rabi_decay_static}
 \zeta(t) = \int \mathrm{d}(\Delta\varepsilon) \,\, N(\Delta\varepsilon) \, \cos \left( \Omega_\mathrm{R} \,t + \phi \right) .
\end{equation}
Under the approximation $\Delta\omega\ll\Omega$,
the Fresnel-type integral (equation~\ref{eq:Rabi_decay_static}) becomes~\cite{Ithier-PRB-05}
\begin{equation} \label{eq:Rabi_decay_static_2}
\zeta(t) = \left(1 + (ut)^2 \right)^{-1/4} \,\cos\left( \Omega \, t + \frac{1}{2}\arctan(ut) \right) ,
\end{equation}
where
\begin{equation} \label{eq:Rabi_decay_static_envelope}
u = \left( \frac{\partial\omega_{01}}{\partial\varepsilon} \right)^2 \frac{\sigma_\varepsilon^2}{\Omega} = \left(\frac{\varepsilon}{\omega_{01}}\right)^2 \, \frac{\sigma_\varepsilon^2}{\Omega} .
\end{equation}

\paragraph{Noise at the Rabi frequency.}
We now turn to the exponential decay of the oscillation envelope ($\Gamma_2$ in the rotating frame),
\begin{equation} \label{eq:Gamma_R}
 \Gamma_\mathrm{R} = \frac{3}{4}\Gamma_1 \cos^2\theta + \frac{1}{2}\Gamma_\Omega ,
\end{equation}
where the component that depends on the noise at the Rabi frequency $\Omega_\mathrm{R}$ has two contributions which we denote $\Gamma_\Omega^{(z)}$ and $\Gamma_\Omega^{(\perp)}$:
\begin{equation} \label{eq:Gamma_Omega}
 \Gamma_\Omega \, = \, \pi S_\varepsilon(\Omega_\mathrm{R}) \, \sin^2\theta \, + \, \pi S_\Delta(\Omega_\mathrm{R}) \, \cos^2\theta \, \equiv \, \Gamma_\Omega^{(z)} \,\sin^2\theta \, + \, \Gamma_\Omega^{(\perp)} \, \cos^2\theta .
\end{equation}
We can approximate $\cos^2\theta \approx 1$, for $\varepsilon\ll\Delta$, and write equation~(\ref{eq:Gamma_R}) as
\begin{equation} \label{eq:Gamma_R_simplified_1}
 \Gamma_\mathrm{R} = \frac{3}{4}\Gamma_1 + \frac{1}{2}\Gamma_\Omega^{(\perp)} + \left(\frac{\varepsilon}{\omega_{01}}\right)^2 \, \frac{1}{2}\Gamma_\Omega^{(z)} .
\end{equation}

\paragraph{Fitting procedure.}
In order to determine $S(\Omega_\mathrm{R})$, we measured the Rabi oscillations vs.\@ $\Phi_\mathrm{b}$, with $\Omega_\mathrm{R}$ fixed for each set of data.
The combined decay envelope from Eqs~(\ref{eq:Rabi_decay_static_2}, \ref{eq:Gamma_R_simplified_1}) becomes
$\left(1 + (ut)^2 \right)^{-1/4}\times$ $\exp(-\Gamma_\mathrm{R} t)$.
We observe that, at $\varepsilon=0$, the model predicts an exponential decay with the rate
$\frac{3}{4}\Gamma_1 + \frac{1}{2}\Gamma_\Omega^{(\perp)}$.
For a given $\Omega_\mathrm{R}$, we find this rate by fitting, and then keep it fixed while fitting the Rabi envelopes vs.\@ $\varepsilon$.
At low amplitude ($\Omega_\mathrm{R}/2\pi\sim2$\,MHz), $\Gamma_\mathrm{R}$ nearly reaches its limit $(3/4)\Gamma_1$. 
Dividing out the known quasi-static contribution, and then fitting the envelope to the parabolic equation~(\ref{eq:Gamma_R_simplified_1}), we obtain the flux-noise dependent $\Gamma_\Omega^{(z)}$, from which we can calculate $S_\varepsilon(\Omega)$.




\subsection*{Energy relaxation; high-frequency spectroscopy.}

\begin{figure}[h!]
\begin{center}
\includegraphics[width=7cm]{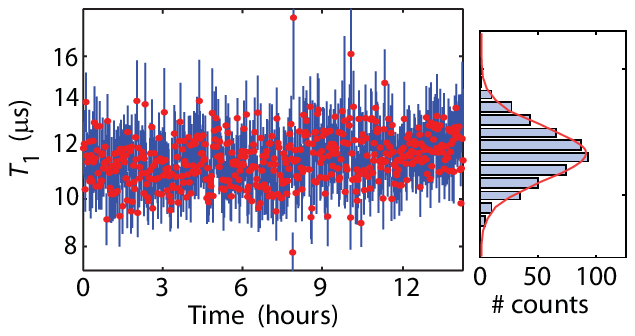}
\begin{quote}
\caption{\textbf{$T_1$ relaxation.}
    Repeated measurements of $T_1$ relaxation from the excited state, at flux degeneracy $\Phi_\mathrm{b}=0$, showing how $T_1$ varies by about 10\,\% from one measurement to the next.
    The size of the error bars from the exponential fits are on the order of the variation.
    Each trace took 1 min 40 s to measure; $T_1$ may vary on a shorter time scale than that.
    }    \label{fig:T1_repeated}
\end{quote}
\end{center}
\vspace{-1cm}
\end{figure}

Figure~\ref{fig:T1_repeated} shows the variation over repeated measurements of $T_1$ relaxation.
To explain the high-frequency noise, leading to $T_1$ relaxation, we model the $\varepsilon$ and $\Delta$ channels of the environment as ohmic resistors.
The microwave antenna's designed mutual inductance to the qubit is
$M=0.02$\,pH with $R=50\,\Omega$ termination.
The Johnson--Nyquist (thermal and quantum) noise of an environment at $T=50$\,mK (a reasonable effective electronic temperature on the chip) is\footnote{Equation~(\ref{eq:J-N_flux}) is written with the following (``type 2") definition of the Fourier transform, used throughout this paper:\\$S_\lambda(\omega) = (1/2\pi) \int_{-\infty}^\infty \mathrm{d}t \, \langle \lambda(0)\lambda(t) \rangle \exp(-i\omega t)$, where $\langle\cdot\rangle$ is the \emph{un-symmetrized} correlation function.}
\begin{equation}
  S^\mathrm{JN}_\varepsilon(\omega) =  \frac{1}{2\pi} \left(\frac{\partial\varepsilon}{\partial\Phi}\right)^2
  M^2  \frac{2 \hbar \omega / R}{1 - e^{-\hbar \omega / k_{\mathrm{B}} T}} ,
\end{equation}
which simplifies to
\begin{equation} \label{eq:J-N_flux}
  S^\mathrm{JN}_\varepsilon(\omega) \approx \frac{1}{2\pi} \left(\frac{\partial\varepsilon}{\partial\Phi}\right)^2 \, M^2  \, \left( 2k_\mathrm{B}T +  2\hbar\omega \right) / R .
\end{equation}
%
This expected noise falls about 100 times below the measured PSD at $f_{01}=\Delta/2\pi$,
indicating that intrinsic, microscopic noise mechanisms dominate the flux noise.
With $R$ unchanged, we would infer an effectively $\sim\!10$ times larger $M$ to account for the observed noise.

Within this work, we did not identify which noise sources were responsible for the $\Delta$ noise, which can arise from, \emph{e.g.\@}, critical-current noise or charge noise due to various microscopic mechanisms.


Taking the tilted quantization axis into account, so that the transverse noise is constituted by $\varepsilon$ noise at $\varepsilon=0$, by $\Delta$ noise at $\varepsilon\gg\Delta$, and by combined projections of the two noises in-between, we arrive at the combined expression for the effective transverse Johnson--Nyquist noise,
\begin{equation}
 S_\lambda(\omega_{01}) = \frac{\Delta^2}{\omega_{01}^2} S^\mathrm{JN}_\varepsilon(\omega_{01}) + \frac{\omega_{01}^2-\Delta^2}{\omega_{01}^2}    S^\mathrm{JN}_\Delta(\omega_{01}) .
\end{equation}
As mentioned above, this \emph{effective} noise is obviously greater than the expected extrinsic noise sources (Johnson--Nyquist as well as $1/f$ noise).
The subgap conductance, typically several hundred to several thousand times the normal resistance, may limit the achievable $T_1$-relaxation times~\cite{Greibe-10}, a topic worth further studies.


\subsection*{Acknowledgements}
We gratefully acknowledge T.~Orlando for support in all aspects of this work.
We appreciate J.~Clarke, L.~Levitov and S.~Lloyd for helpful discussions, and M.~Biercuk, P.~Forn-Diaz and S.~Valenzuela for comments on the manuscript.
We thank P.~Murphy and the LTSE team at MIT Lincoln Laboratory for technical assistance.
This work was sponsored by the U.S. Government; the Laboratory for Physical Sciences; the National Science Foundation;
and the Funding Program for World-Leading Innovative R\&D
on Science and Technology (FIRST), CREST-JST, MEXT kakenhi ``Quantum Cybernetics".

Opinions, interpretations, conclusions and recommendations are those of the author(s) and are not necessarily endorsed by the U.S.\@ Government.


\end{document}